\documentclass[aps,11pt,prd,groupedaddress,nofootinbib,notitlepage,eqsecnum,preprintnumbers]{revtex4-2}

\usepackage[utf8]{inputenc}
\usepackage{graphicx}
\usepackage{amsmath, amssymb}
\usepackage{mathtools}
\usepackage{physics}
\usepackage[english]{babel}
\usepackage[colorlinks=true, allcolors=blue]{hyperref}
\usepackage{url}
\usepackage{hyperref}
\usepackage{cleveref}
\usepackage{xcolor}
\usepackage[normalem]{ulem}
\usepackage{soul}
\usepackage{bbold}
\usepackage[lofdepth,lotdepth,caption=false]{subfig}

\graphicspath{{./figures/}}

\def\calH{\mathcal{H}}
\def\Mpl{M_{\rm Pl}}

\definecolor{elpurple}{HTML}{a300ff}

\begin{document}

\title{Gravitational waves induced by matter isocurvature in general cosmologies}

\author{\textsc{Guillem Dom\`enech$^{a,b}$}}
    \email{{guillem.domenech}@{itp.uni-hannover.de}}
\author{\textsc{Jan Tränkle$^{a}$}}
\email{{jan.traenkle}@{itp.uni-hannover.de}}

\affiliation{$^a$Institute for Theoretical Physics, Leibniz University Hannover, Appelstraße 2, 30167 Hannover, Germany}
\affiliation{$^b$ Max-Planck-Institut für Gravitationsphysik, Albert-Einstein-Institut, 30167 Hannover, Germany}

\begin{abstract}
    The expansion history and content of the Universe between the end of inflation and the onset of Big Bang Nucleosynthesis is mostly unknown. In this paper, we study gravitational waves (GWs) induced by matter isocurvature fluctuations in a generic perfect fluid background as a novel probe of the physics of the very early Universe. We analytically compute the induced GW kernel and analyze the spectral GW energy density for a sharply peaked isocurvature power spectrum. We show that the spectral shape of the GW signal is sensitive to the equation of state parameter $w$ of the perfect fluid dominating the early Universe after inflation. We find that the GW amplitude is enhanced for a soft equation of state. Our framework can be applied to dark matter isocurvature and models leading to early matter-dominated eras, such as primordial black holes and cosmological solitons.
\end{abstract}

\maketitle

\section{Introduction} \label{sec:Introduction}
Observations of the cosmic microwave background (CMB) \cite{Planck:2018jri} have led to the paradigm of cosmic inflation \cite{Starobinsky:1979ty, Sato:1981qmu, Guth:1980zm, Starobinsky:1980te, Kazanas:1980tx, Linde:1981mu, Albrecht:1982wi} as the leading model for the very early Universe, while the study of Big Bang Nucleosynthesis (BBN) \cite{Grohs:2023voo} has revealed that the subsequent radiation-dominated epoch began at the latest when the Universe had cooled to a temperature of a few MeV \cite{Kawasaki:1999na,Kawasaki:2000en,Hannestad:2004px,Hasegawa:2019jsa}. The transition between these two periods is referred to as reheating \cite{Allahverdi:2010xz, Amin:2014eta} and remains mostly unconstrained, see e.g.~\cite{Allahverdi:2020bys} for a review of possible expansion histories.

Typical models of reheating involve a period of early matter domination (eMD), where the average equation of state (EoS) parameter $w$ is dust-like, $w \simeq 0$. This may, for example, be the case if the Universe is temporarily dominated by a coherently oscillating superheavy scalar field (e.g.~string-theoretic moduli fields), or if there is a period where discrete, non-relativistic objects, such as primordial black holes (PBHs), solitons, Q-balls or oscillons come to dominate the energy density of the Universe \cite{Turner:1983he, Drees:2017iod, Kane:2015qea, Amin:2011hj, Allahverdi:2020bys, Khlopov:2008qy, Escriva:2022duf, Byrnes:2025tji}.
However, in general, the EoS parameter during reheating could be anywhere between $-1/3< w \leq 1$, depending on the shape of the inflaton potential and the details of the reheating dynamics \cite{Dodelson:2003ft}. For example, any value between $0\leq w\leq 1$ can be achieved, if the inflaton oscillates coherently in a polynomial potential $V(\varphi)\propto |\varphi|^{2n}$ with generic $n$, or in an exponential potential of the form $V(\varphi)\propto e^{-\lambda \varphi}$ (the latter also leads to arbitrary values for $w$). In these cases, the effective equation of state parameter is determined by the parameters of the potential as $w={(n-1)}/{(n+1)}$ \cite{Turner:1983he} and $w=\lambda^2 /3 -1$ \cite{Lucchin:1984yf}, respectively.
One particularly interesting scenario involves a period of kination, where $w\approx 1$. Such a period often arises in quintessential inflation scenarios, where the inflaton accelerates in a runaway potential after inflation \cite{Spokoiny:1993kt, Joyce:1996cp, Peebles:1998qn, WaliHossain:2014usl}. Moreover, scalar condensates from monomial potentials, i.e.~$V(\varphi)\propto |\varphi|^{2n}$, have recently received interest with respect to dark matter \cite{Garcia:2020eof,Clery:2024dlk, Haque:2021mab, Haque:2023yra} and gravitational wave (GW) production \cite{Bernal:2023wus,Choi:2024ilx,Gross:2024wkl, Haque:2021dha}.

One promising avenue to probe reheating expansion histories is via the generation of induced GWs, see e.g.~\cite{Bhattacharya:2019bvk, Domenech:2019quo, Domenech:2020kqm, Dalianis:2020cla, Witkowski:2021raz, Domenech:2024wao, Bhaumik:2024qzd, Paul:2025kdd, Maiti:2024nhv, Chakraborty:2024rgl, Maiti:2025cbi} for studies involving general $w$ and \cite{Assadullahi:2009nf, Jedamzik:2010hq, Alabidi:2013lya, Inomata:2019zqy, Inomata:2019ivs, Dalianis:2020gup, Pearce:2023kxp, Kumar:2024hsi} for generic eMD eras. For applications to PBH, Q-ball, and oscillon domination, see Refs.~\cite{Inomata:2020lmk, Papanikolaou:2020qtd, Domenech:2020ssp, Domenech:2021wkk, Domenech:2024wao, del-Corral:2025fca, Bhaumik:2024qzd, Paul:2025kdd, Lozanov:2022yoy,Kawasaki:2023rfx}, respectively. In the standard picture, these GWs are generated at second order in cosmological perturbation theory from the backreaction of large, adiabatic scalar fluctuations \cite{Tomita:1967wkp, Matarrese:1992rp, Matarrese:1993zf, Carbone:2004iv, Ananda:2006af, Baumann:2007zm, Kohri:2018awv}, see \cite{Domenech:2021ztg, Yuan:2021qgz, Domenech:2025ior} for recent reviews. However, large isocurvature perturbations can also lead to a sizable production of induced GWs \cite{Domenech:2021and, Domenech:2023jve}. Interestingly, isocurvature perturbations are inevitably generated in scenarios involving localized objects like PBHs, solitons, Q-balls or oscillons, which are of Poisson nature \cite{Meszaros:1975ef, Papanikolaou:2020qtd}.
These Poisson fluctuations lead to a “universal” GW signature from the formation of any cosmological solitons \cite{Lozanov:2023aez, Lozanov:2023knf}. The isocurvature-induced GW signal is largest if the solitons come to dominate the Universe.
Additional fields during inflation may also develop a sizeable isocurvature \cite{Linde:1985yf, Kodama:1986fg, Kodama:1986ud, Polarski:1994rz, Lyth:2001nq, Langlois:2003fq, Kasuya:2009up} and induced GW spectrum, see e.g.~\cite{Bartolo:2007vp, Kumar:2024hsi, Garcia:2025yit}.

It should be noted that isocurvature is strongly constrained on the largest scales via CMB \cite{Langlois:2003fq,Planck:2018jri}, spectral distortion \cite{Chluba:2013dna}, Ly-$\alpha$ observations \cite{Buckley:2025zgh}, and BBN \cite{Inomata:2018htm}. But there are almost no constraints on smaller scales. In fact, large isocurvature perturbations can even lead to the formation of PBHs \cite{Yokoyama:1995ex,Passaglia:2021jla}. Induced GWs offer a new way to test isocurvature fluctuations. In this paper, we generalize earlier works \cite{Domenech:2021and, Lozanov:2023aez, Lozanov:2023knf} by computing isocurvature-induced GWs in general cosmological backgrounds with a constant $0<w\leq 1$. In particular, we extend our previous work \cite{Domenech:2024wao} to include the isocurvature-induced GWs produced before a matter-dominated phase (though our formalism is also valid even if the matter component never dominates). We will show that the resulting isocurvature-induced GW spectrum differs significantly from the previously studied radiation-matter case.
Although $w$ may in general vary over time \cite{Saha:2020bis, Lozanov:2016hid, Carr:2019kxo, Antusch:2025ewc}, our calculations constitute an analytical benchmark case. Nevertheless, if reheating is long enough, there will be a period with an almost constant $w$.

The paper is structured as follows.
In \cref{sec:Isocurvature} we begin by discussing the evolution of curvature perturbations sourced by an initial isocurvature fluctuation.
The analytical approximation we derive will allow us to compute the induced GW kernel in \cref{sec:Induced_GWs}. After deriving the induced GW source term in \cref{sec:Induced_GWs_Source}, we discuss the computation of the kernel in some detail in \cref{sec:Induced_GWs_Kernel}.
Using this result, we compute the GW spectral energy density in \cref{sec:Induced_GW_spectrum}. We focus on a sharply peaked isocurvature spectrum and discuss features of the spectrum, with a particular emphasis on the dependence on the equation of state parameter and the different contributions.
\Cref{sec:Summary} contains a summary of our results and gives some conclusions.

\section{Matter isocurvature perturbations} \label{sec:Isocurvature}
We start by reviewing the evolution of matter isocurvature fluctuations in a general cosmological background as derived in Ref.~\cite{Domenech:2024wao}. First, we consider a Friedmann–Lemaître–Robertson–Walker (FLRW) universe filled with two components, one of which behaves like non-relativistic matter (or dust) and the other can be treated as an adiabatic, perfect fluid with an equation of state $w=P/\rho$, defined as the ratio of the fluid's pressure and energy density, with $w$ in the range $0<w\leq 1$.
We assume that the matter component is initially subdominant, meaning $\rho_{ \text{m},i}/\rho_{w,i} \ll 1$ for the energy densities at the initial time, denoted by the subscript “$i$”. As the universe expands, the matter component will eventually start dominating the background because its energy density redshifts as $\rho_m\propto a^{-3}$ with the scale factor $a$, compared to the primordial fluid, which redshifts faster with $\rho_w \propto a^{-3(1+w)}$ since $w>0$ by assumption. We denote the time, when the two energy densities become equal, $\rho_{w, w\text{eq}}=\rho_{m,w\text{eq}}$, with the subscript “$w$eq” (for “matter-$w$ equality”, not to be confused with the later (cold dark) matter-radiation equality).

To study the evolution of fluctuations about the homogeneous background, we adopt a perturbed flat FLRW metric in the conformal Newtonian (also called shear-free, or longitudinal) gauge,
\begin{equation} \label{eq:FLRW}
    ds^2 = a^2(\tau) \left[ - (1-2 \Phi)\text{d}\tau^2 + \left( \delta_{ij} + 2 \Phi \delta_{ij} + h_{ij}\right) \text{d}x^i \text{d}x^j \right] \,,
\end{equation}
with the Newtonian gravitational potential $\Phi$ (i.e.~the Bardeen potential in the Newton gauge), and the transverse-traceless tensor perturbations $h_{ij}$, which on subhorizon scales constitute gravitational waves. We assume that anisotropic stress is negligible, such that the scalar sector is fully determined by the single potential $\Phi$, and we neglect any vector perturbations. Conformal time $\tau$ is defined by $\text{d}t=a\text{d}\tau$ in terms of the scale factor $a$ and cosmic time $t$. For more details on the theory of cosmological perturbations, we refer the reader to Refs.~\cite{Dodelson:2003ft, Malik:2008im, Baumann:2022mni}. We provide details on the resulting Einstein and fluid equations in \cref{app:einsteinequations}.

One can obtain the equation of motion for $\Phi$ by combining the trace of the $ij$-component and the $00$-component of the perturbed Einstein equations, \cref{eq:ijEinsteinEq,eq:00EinsteinEq}, respectively. One finds
\begin{equation}
    \Phi'' +  3 \calH(1+c_s^2)\Phi' + \left((1+3c_s^2)\calH^2+2\calH' +c_s^2 k^2\right)\Phi = \frac{1}{2}a^2c_s^2\rho_{\rm m} S \,,
    \label{eq:PhiEq}
\end{equation}
where $'\equiv d/d\tau$, $\calH = a'/a$ is the conformal Hubble parameter, and $k$ denotes the wavenumber of a given mode in Fourier space. The sound speed $c_s^2$ is defined in \cref{eq:sound_speed} and interpolates between $c_s^2\approx w$ at early times when $\rho_m\ll \rho_w$, and $c_s^2\approx 0$ at late times. Here, for simplicity, we have assumed that the non-adiabatic pressure of the primordial fluid vanishes, such that its sound speed is given by $c_w^2=w$.
In \cref{eq:PhiEq} we introduced the isocurvature perturbation
\begin{equation} \label{eq:S_Def}
    S\coloneq \frac{\delta\rho_{\rm m}}{\rho_{\rm m}}-\frac{\delta\rho_w}{(1+w)\rho_w} \,,
\end{equation}
in terms of the perturbations of the energy densities $\delta\rho_{m/w}$ of the two components.
The evolution of $S$ is governed by \cref{eq:SEq}, which can be derived from the energy and momentum conservation equations \labelcref{eq:Energy_Cons_m,eq:Energy_Cons_w,eq:Momentum_Cons_m,eq:Momentum_Cons_w}.
If there is a non-vanishing isocurvature fluctuation $S_i\neq 0$ at early times, it will, through \cref{eq:PhiEq}, source curvature perturbations as the universe evolves, even if the initial curvature perturbation vanishes, $\Phi_i =0$. This induced curvature perturbation, then, at second order in cosmological perturbations, sources tensor perturbations, i.e.~GWs, which would potentially allow us to probe such small-scale fluctuations.

The coupled system of equations for the gravitational potential $\Phi$ and the isocurvature perturbation $S$, \cref{eq:PhiEq,eq:SEq}, can be solved analytically deep inside the early, $w$-dominated epoch.
Following \cite{Domenech:2021and,Domenech:2024wao}, we expand the equations for small scale factor $a \ll a_{w \rm eq}$ and choose a perturbative ansatz for $\Phi$ and $S$. Note that this expansion becomes a poor approximation when $w\ll 1$ as the ratio of energy densities approaches a constant. For instance, in the exact $w\to0$ limit, the right-hand side of \cref{eq:PhiEq} trivially vanishes and no curvature perturbation is sourced. For the moment, we work in this perturbative expansion and evaluate the validity of the approximation later. 
Now, since $S$ is constant in time on superhorizon scales, we can at first order treat ${S(k\ll \calH) \approx S_i}$ as a constant source for $\Phi$. Note that $S_i\equiv S_{i,\textbf{k}}$ is drawn from some initial distribution specified by its power spectrum $\mathcal{P}_{S}(k) \delta_D(k+k')=\frac{k^3}{2\pi^2}\langle S_{i,\textbf{k}} S_{i,\textbf{k}'} \rangle$, and is therefore generally scale-dependent. Using the Green's function method, the solution to \cref{eq:PhiEq} can then formally be written as \cite{Domenech:2024wao}
\begin{align}
    \Phi_{\rm iso}(x)= &-2^{\frac{b-5}{2}}\pi (b-1) (b+1)^b \sec (\pi  b) \kappa ^{b-1} x^{-b-\frac{3}{2}}  S_i\nonumber \\
    &\times\left( J_{b+\frac{3}{2}}\left(c_s x\right) \int_0^x d\tilde{x} \, J_{-b-\frac{3}{2}}\left(c_s \tilde{x}\right) \tilde{x}^{3/2}
    - J_{-b-\frac{3}{2}}\left(c_s x\right) \int_0^x d\tilde{x} \, J_{b+\frac{3}{2}}\left(c_s \tilde{x}\right) \tilde{x}^{3/2} \right) \,,
    \label{eq:Phi_iso_Green_sol}
\end{align}
where $J_\nu$ denotes the Bessel functions of the first kind, and we defined for compactness
\begin{equation}
    b\coloneq\frac{1-3w}{1+3w}\,.
    \label{eq:bDefinition}
\end{equation}
The value of $b$ ranges from $b=-1/2$ for $w=1$ (corresponding to a kination-like period), over $b=0$ for early radiation domination ($w=1/3$), to $b=1$ for $w=0$. We refer to values of $w<1/3$ ($b>0$) as soft, while a stiff EoS means $w>1/3$ ($b<0$). In \cref{eq:Phi_iso_Green_sol}, the time variable is chosen as $x\coloneq k \tau$, which defines the superhorizon ($x\ll1$) and subhorizon ($x\gg1$) regimes for a given mode $k$. We also introduced the parameter 
\begin{align}
\kappa \coloneq k/k_{w \rm eq}\,,
\end{align} 
which measures the ratio of the wavelength of a perturbation mode and the size of the (comoving) Hubble radius at $w$-matter equality, $\calH_{w \rm eq}=k_{w \rm eq}$. We will be interested in small-scale modes, $\kappa \gg 1$, which re-enter the cosmological horizon deep inside the $w$-dominated epoch. More details on the derivation and the full expression for \cref{eq:Phi_iso_Green_sol} are provided in \cite{Domenech:2024wao}.

Performing the integrals in \cref{eq:Phi_iso_Green_sol} results in expressions involving generalized hypergeometric functions $_1 F_2$, inconvenient for later integration.
For practical purposes, we find an excellent approximation to \cref{eq:Phi_iso_Green_sol} as follows. First, we note that \cref{eq:Phi_iso_Green_sol} has the following asymptotic behaviors, namely
\begin{align}
    \Phi_{\rm iso}(x) & \propto S_i \, \kappa^{-1+b}  x^{-(1+b)}\times
    \begin{cases}
        x^2 & \quad (x\ll 1)\\
       1 + x^{-1} \sin(c_sx) & \quad (x\gg 1)
    \end{cases} \,, \label{eq:Phi_iso_scaling}
\end{align}
where $c_s\approx\sqrt{w}$ and we neglected constant prefactors.
From \cref{eq:Phi_iso_scaling}, we see that the isocurvature-induced curvature perturbation behaves as in the radiation-matter case \cite{Kodama:1986ud,Domenech:2021and}, except for an overall $
(x/\kappa)^{-b}$ prefactor. Moreover, the oscillations in $\Phi$ have the same phase independent of $b$.\footnote{For comparison, adiabatic initial conditions $(\Phi_i\neq0, \, S_i=0)$ yield \cite{Baumann:2007zm, Domenech:2024wao}
\begin{align} 
    \Phi_{\rm ad}(x)  \propto \Phi_i\times
    \begin{cases}
        1 & \quad (x\ll 1)\\
       x^{-2-b} \cos(c_s x - \frac{b \pi}{2}) & \quad (x\gg 1)
    \end{cases} \,. \label{eq:Phi_ad_scaling}
\end{align}
See how the phases of the subhorizon oscillations depend on $b$.} Since $|b|<1$, we can use the radiation-matter solution with $c_s=\sqrt{w}$ and demand the correct asymptotic scaling for large and small $x$. Corrections will be proportional to $b$ and, therefore, are expected to be small. However, we need an additional function to fit both regimes. We find that the spherical Bessel function $j_2(c_s x)$, which happens to be a solution to \cref{eq:PhiEq} for $b=1$, has the desired asymptotic behaviors. With this procedure, we arrive at
\begin{align}
    \Phi_{\rm iso}(x) \approx 2^{\frac{b-3}{2}} 3 (b+1)^{b+1} \kappa ^{b-1} x^{-(1+b)}S_i\left(1+\frac{2}{c_s^2 x^2}+ 2 y_1(c_s x)+\frac{15 b (b+3)}{4\left(4-3 b-b^2\right)}\,j_2(c_s x)\right)  \,,
    \label{eq:Phi_Iso_GD}
\end{align}
where $y_1$ is the spherical Bessel function of the second kind. For $b=0$, \cref{eq:Phi_Iso_GD} recovers the exact solution \cite{Domenech:2021and}.

We checked numerically that the approximation \labelcref{eq:Phi_Iso_GD} agrees with the approximate analytical solution \eqref{eq:Phi_iso_Green_sol} at a precision of better than $\mathcal{O}(10^{-1}-10^{-2})$ in most of the range of interest. It recovers exactly the super- and subhorizon asymptotic behaviors by construction. 
However, as anticipated, for very soft equation of state parameters ($w\ll1$) the approximation leading to \cref{eq:Phi_iso_Green_sol} breaks down, as the transition of the background is very gradual and the energy density of both fluids is comparable also for quite small $a < a_{w \rm eq}$. This manifests itself in the analytical solution for $b\gtrsim 0.65$, when the solution starts changing sign in the oscillations.
This is actually a feature seen for adiabatic initial conditions, but not in the numerical solution with isocurvature ones. This implies that our approximation starts behaving like the adiabatic solution for $b\gtrsim 0.65$ and, in some sense, fails to resolve the two fluids separately and effectively treats them as a single fluid with adiabatic initial conditions. To avoid such artifacts at later stages, we restrict ourselves to $b\lesssim 0.65$ ($w\gtrsim 7 \times 10^{-2}$). If a more accurate solution for $b\approx 1$ is needed, one could approach this issue by expanding the equations of motion \cref{eq:PhiEq,eq:SEq} about $w\approx 0$.
Additionally, for very soft EoS parameters, we require large $\kappa \gg 1$, such that scales do not re-enter the horizon too close to the transition regime.
%
\begin{figure}
\centering
\includegraphics[width=0.6\textwidth]{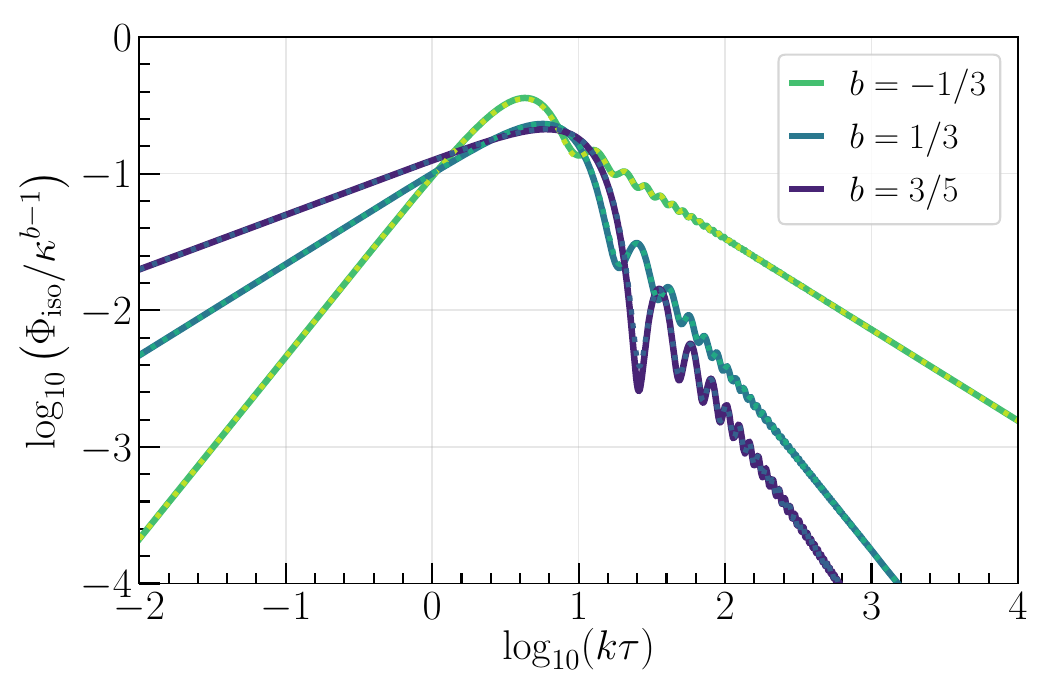}
\caption{The evolution of the gravitational potential $\Phi_{\rm iso}(k\tau)$ with isocurvature initial conditions, normalized by $\kappa^{b-1}$. We show some examples of soft and stiff equations of state ($b=-\frac{1}{3}, \, \frac{1}{3}, \, \frac{3}{5}$, corresponding to $w=\frac{2}{3}, \, \frac{1}{6}, \, \frac{1}{12}$), with fixed $\kappa = 10^5$. Solid lines represent the approximate solution \labelcref{eq:Phi_Iso_GD}, whereas the dotted line, which lies almost perfectly on top of the solid one, shows the exact analytical solution of \cref{eq:Phi_iso_Green_sol}. In \cref{fig:Phi_iso_2} in \cref{app:einsteinequations}, we also include the result of numerical integration.
}
\label{fig:Phi_iso}
\end{figure}
%

In \cref{fig:Phi_iso} we plot the time evolution of the gravitational potential $\Phi_{\rm iso}(x)$ for $\kappa = 10^5$ and some representative values of $w$, illustrating the effect of a stiffer or softer EoS. See how the slope in the super- and subhorizon regimes changes with $w$, and how after horizon entry at $k\tau \sim 1$ the potential starts oscillating and decaying. It is around this time that most gravitational waves are sourced by the scalar perturbation.
We normalized $\Phi_{\rm iso}$ by the factor $\kappa^{b-1}$, which would lead to a relative suppression of the amplitude of the curvature perturbation for a stiffer equation of state due to the larger pressure.
Note that our approximation ceases to be valid close to and beyond the transition from $w$- to matter domination, and for very soft EoS $b\gtrsim 0.65$. We illustrate the regime of validity of the approximation in \cref{fig:Phi_iso_2} in \cref{app:einsteinequations}.

In passing, let us highlight the key differences in the behavior of the curvature perturbation for isocurvature and adiabatic initial conditions (see \cref{eq:Phi_iso_scaling,eq:Phi_ad_scaling}, respectively). Firstly, we see that the amplitude of the isocurvature-induced curvature perturbation is suppressed by a factor $\kappa^{-1+b}$, which reflects the fact that for modes that enter the horizon early, the isocurvature has not been fully converted into curvature yet, before they start decaying due to the non-zero pressure.
We also note that the two solutions oscillate out of phase, with exactly opposite phases in the radiation case $b=0$ \cite{Domenech:2021and, Domenech:2023jve}.
Furthermore, we see that in the adiabatic case the leading order term is a damped oscillation, meaning that the curvature perturbation changes sign as it oscillates, while in the isocurvature case the leading order is just a power-law, with damped oscillations appearing in the first subleading correction, such that $\Phi_{\rm iso}$ does not change sign.
Finally, we see that in the limit $b\rightarrow 1$, where both fluids have the same equation of state, the scaling of the isocurvature solution approaches that of the adiabatic case, because in this limit, there is no relative evolution between the two components and fluctuations behave essentially adiabatically. Recall, however, that for $b\gtrsim0.65$ our analytical approximation ceases to be valid.

\section{Induced gravitational waves} \label{sec:Induced_GWs}
We now turn to study the gravitational waves generated by the isocurvature-induced curvature perturbation \labelcref{eq:Phi_Iso_GD} around the time of horizon entry.
For a review on induced GWs in general, we refer to \cite{Domenech:2021ztg}, and for a review on isocurvature-induced GWs in a radiation background, see \cite{Domenech:2023jve}. For a discussion on the gauge dependence of isocurvature-induced GWs, see \cite{Yuan:2024qfz}. Here, we follow closely the approach of Ref.~\cite{Domenech:2021and}. We first review the derivation of the source term and later compute the induced GW kernel.

\subsection{Induced GW source term} \label{sec:Induced_GWs_Source} 
The equations of motion for the tensor perturbations at second order are obtained from the transverse-traceless part of the $ij$-component of the perturbed Einstein equations, and read
\begin{align}
    h_{ij}'' + 2 \calH h_{ij}' - \partial_a\partial^a h_{ij} = \mathcal{P}_{ij}^{\text{TT}, ab}\mathcal{S}_{ab} \,,
    \label{eq:Tensor_EoM}
\end{align}
where we defined the source term
\begin{align}
   \mathcal{S}_{ab} = 4\partial_a \Phi \partial_b \Phi + 2 a^2\left((1+w)\rho_w \partial_a V_w \partial_b V_w + \rho_{\rm m} \partial_a V_{\rm m} \partial_b V_{\rm m} \right) \, ,
   \label{eq:source_term_Phi_V}
\end{align}
in terms of the gravitational potential $\Phi$, the fluid velocities $V_{\rm m}$ and $V_w$, and $\mathcal{P}_{ij}^{\text{TT}, ab}$ is the transverse-traceless projection operator \cite{Domenech:2021ztg}. The first term $\sim(\partial \Phi)^2$ originates from the perturbed Einstein tensor $G_{\mu\nu}$ and can be understood as a consequence of the non-linear nature of gravity coupling scalar and tensor modes at second order, while the remaining terms containing the velocities stem from the energy-momentum tensor $T_{\mu\nu}$ of the matter content.
\Cref{eq:source_term_Phi_V} illustrates that large velocity flows in the matter sector, as well as gradients in the gravitational potential, can source an induced GW signal.

Defining the total and relative velocities for the two components as
\begin{align}
    V \coloneq \frac{\rho_{\rm m} V_{\rm m} +(1+w) \rho_w V_w}{\rho_{\rm m} + \rho_w} \quad \text{and} \quad V_{\rm rel} \coloneq V_{\rm m} - V_w \,,
\end{align}
we can rewrite the source term as
\begin{align}
    \mathcal{S}_{ab} & = 4\partial_a \Phi \partial_b \Phi 
    + \frac{2a^2 \rho^2}{\rho_{\rm m} + (1+w) \rho_w} \left(\partial_a V \partial_b V + (1+w) \frac{\rho_{\rm m} \rho_w}{\rho^2} \partial_a V_{\rm rel} \partial_b V_{\rm rel} \right) \\
    & = 4\partial_a \Phi \partial_b \Phi 
    + \frac{8}{3}\frac{\rho}{(1+w) \rho_w} \frac{c_s^2}{c_w^2} \partial_a \left(\Phi'/\calH +\Phi \right)
    \partial_b \left(\Phi'/\calH +\Phi \right) 
    + 2 a^2 \rho_{\rm m} \frac{c_s^2}{c_w^2} \partial_a V_{\rm rel} \partial_b V_{\rm rel} \,,
    \label{eq:source_term_S_ab}
\end{align}
where for the second equality we used the Friedmann equation \labelcref{eq:FriedmannEq} and the $0i$-component of the perturbed Einstein equations \labelcref{eq:0iEinsteinEq}.
One sees that at early times, i.e., for $\rho_w \gg \rho_{\rm m}$ (or $a\ll a_{w \rm eq}$) where $c_s \simeq c_w$, the last term containing the relative velocity $V_{\rm rel}$ is suppressed by a factor $\rho_{\rm m}$.\footnote{Relative velocities may become relevant close to the time of equality \cite{Domenech:2020ssp,Kumar:2024hsi}.}

In the limit $\rho_w \gg \rho_{\rm m}$ deep during $w$-domination, at leading order we obtain
\begin{align}
    \mathcal{S}_{ab}(a\ll a_{w \rm eq}) & \approx 4\partial_a \Phi \partial_b \Phi 
    + 4\frac{1+b}{2+b} \partial_a \left(\Phi'/\calH +\Phi \right)
    \partial_b \left(\Phi'/\calH +\Phi \right) \,,
    \label{eq:source_term_S_ab_wDE}
\end{align}
such that the source term is expressed purely in terms of the gravitational potential $\Phi$ and its time derivative. 
We will now turn to computing the GW spectrum generated by the time evolution of the isocurvature-induced curvature perturbation $\Phi$ as given in \labelcref{eq:Phi_Iso_GD}.
Later, we will find that the contribution due to the total velocity perturbation $V$, i.e., the terms containing time derivatives of $\Phi$, is the dominant one.

\subsection{Computation of the kernel} \label{sec:Induced_GWs_Kernel} 
The tensor equation of motion \labelcref{eq:Tensor_EoM} can formally be solved with the Green's function method. To this end, we define the kernel for the induced GWs, carrying the full time-dependence, as
\begin{equation}
    I(x,k,u,v)=\int_{x_i}^x d\tilde{x} \ \mathcal{G}_h(x,\tilde{x}) f(\tilde{x},k,u,v) \,,
    \label{eq:Kernel_I}
\end{equation}
where the tensor modes' Green's function is given in \cref{eq:TensorGreensFctGen} and the source term $f(x,k,u,v)$, provided explicitly in \cref{eq:kernel_fxuv_expr}, results directly from \cref{eq:source_term_S_ab_wDE}.
\Cref{eq:Kernel_I} encodes that a tensor mode $h_\textbf{k}$ with momentum $k$ is sourced by two scalar modes with momenta $u k$ and $v k$, respectively. The integral over the Green's function times the source captures the response of the tensors to the presence of the source from time $x_i$ until $x$. 
After substituting the expression \cref{eq:TensorGreensFctGen} for the Green's function, we can split the integral in \cref{eq:Kernel_I} as
\begin{align}
    I(x,k,u,v) = &  x^{-b-\frac{1}{2}} \kappa^{2 b-2} \left(Y_{b+\frac{1}{2}}(x) \mathcal{I}_J(u,v)+ J_{b+\frac{1}{2}}(x) \mathcal{I}_Y(u,v) \right) \,,
    \label{eq:KernelI_split}
\end{align}
where we pulled out the $k$-dependence of the source as $f(x,k,u,v)=\kappa^{2 b-2} \tilde{f}(x,u,v)$, and defined
\begin{equation}
    \mathcal{I}_{J/Y} \left(u,v\right) = \frac{\pi}{2}\int_{x_i}^x d\tilde{x} \, \tilde{x}^{b+3/2}
    \begin{rcases}
        \begin{dcases}
            J_{b+\frac{1}{2}}(\tilde{x}) \\
            Y_{b+\frac{1}{2}}(\tilde{x})
        \end{dcases}
    \end{rcases}
    \tilde{f}(\tilde{x},u,v) \,.
    \label{eq:Kernel_I_JY_definition}
\end{equation}

Conceptually, at late times ($x\gg 1$) one may think of the $x$-dependent Bessel functions in \cref{eq:KernelI_split} as the oscillating tensor modes (i.e.~the GWs), with the $\mathcal{I}$-terms determining their amplitude.
Unfortunately, we have not found analytical expressions for the integrals in \cref{eq:Kernel_I_JY_definition} for general $b$. We thus proceed with the following approximation. We split the integrals over the time coordinate $\tilde{x}$ in the full kernel into super- and subhorizon parts ($\tilde{x}\ll 1$ and $\tilde{x}\gg 1$, respectively) as
\begin{align}
    \mathcal{I}_{J} \left(u,v\right) \approx \int_{x_i}^{\xi_J} d\tilde{x} \ \tilde{x}^{b+3/2} J_{b+\frac{1}{2}}(\tilde{x} \ll 1) \tilde{f}(\tilde{x},u,v) + \int_{\xi_J}^{x} d\tilde{x} \ \tilde{x}^{b+3/2} J_{b+\frac{1}{2}}(\tilde{x} \gg 1) \tilde{f}(\tilde{x},u,v) \,,
    \label{eq:KernelI_sub_super_split}
\end{align}
and we denote with $\mathcal{I}_J^{\tilde{x}\ll 1}(u,v)$ and $\mathcal{I}_J^{\tilde{x}\gg 1}(u,v)$ the first and second term in \cref{eq:KernelI_sub_super_split}, respectively.
For $\mathcal{I}_{Y} \left(u,v\right)$ we proceed analogously.
The asymptotic expansions of the Bessel functions for small and large argument are given in \cref{eq:BesselJ_Smallx,eq:BesselJ_Largex,eq:BesselY_Smallx,eq:BesselY_Largex}, respectively.
The integration boundary $\xi_{J/Y}$ is an $\mathcal{O}(1)$ matching point, and we define it in \cref{eq:matching_points} for the two cases.
We will further take $x_i\rightarrow 0$ and $x\rightarrow \infty$ to simplify the evaluation of the integrals, which is justified as the source term $f(x,u,v)$ goes to zero in both limits, and we are interested in the tensor power spectrum at late times, when the tensor modes are deep inside the horizon and behave as GWs. This approximation also requires $\kappa\gg1$, such that the generation of GWs is completed before the transition of the background.

In the following sections, we will outline our strategy for analytically solving the time integrals in \cref{eq:KernelI_sub_super_split} and give the structure of the solutions. In \cref{app:derivation_kernel}, we provide some more details on the calculation.
The computations were performed with \texttt{Mathematica}, and we provide the resulting lengthy expressions for the coefficients in \cref{app:coefficients}.

\subsubsection{Superhorizon integrals}
Let us start by considering the superhorizon ($\tilde{x}\ll 1$) integrals.
We begin by expanding the integrands and reducing all products of trigonometric functions to single sine and cosine terms, using the identities given in \cref{eq:trig_identities}, such that the integrals reduce to a sum of integrals of powers of $\tilde{x}$ times elementary trigonometric functions. Further, we perform partial integrations of the form \cref{eq:partial_int} to reduce some of the higher powers. The schematic form of the resulting integrals is given in \cref{eq:KernelIJ_super_integral}.
In this form, the integrals can readily be performed, and one can write the result as a sum of trigonometric and hypergeometric functions with different arguments.

For compactness, we define the generalized sine and cosine integrals by
\begin{align}
    \text{si}(n,m,z) & \coloneq \int_z^\infty dx \, x^{-n} \sin(m x) \nonumber \\
    &= \mathfrak{si}(n,m,z) + \frac{| m| ^n}{m} \cos \left(\frac{\pi  n}{2}\right) \Gamma (1-n) \,, \label{eq:si(nmz)} \\
    \text{ci}(n,m,z) & \coloneq \int_z^\infty dx \, x^{-n} \cos(m x) \nonumber \\
    &= \mathfrak{ci}(n,m,z)+|m|^{n-1} \sin \left(\frac{\pi  n}{2}\right) \Gamma (1-n) \,, \label{eq:ci(nmz)}
\end{align}
which are well-defined for $z>0$ and $n>0$, and where
\begin{align}
    \label{eq:si(nmz)_reg}
    \mathfrak{si}(n,m,z) &\coloneq \frac{m z^{2-n}}{n-2} \,_1F_2\left(1-\frac{n}{2};\frac{3}{2},2-\frac{n}{2};-\frac{1}{4} m^2 z^2\right) \,,\\
    \label{eq:ci(nmz)_reg}
    \mathfrak{ci}(n,m,z) &\coloneq \frac{z^{1-n}}{n-1} \,_1F_2\left(\frac{1}{2}-\frac{n}{2};\frac{1}{2},\frac{3}{2}-\frac{n}{2};-\frac{1}{4} m^2 z^2\right) \,.
\end{align}
The definition of the normalized versions $\mathfrak{si}(n,m,z)$ and $\mathfrak{ci}(n,m,z)$ of the generalized sine/cosine integrals turns out to be convenient, as the constant terms in \cref{eq:si(nmz),eq:ci(nmz)} drop out in the final result, as can be seen from \cref{eq:split_integral}.
The standard sine and cosine integrals, Si$(z)$ and Ci$(z)$, in this notation are given by
\begin{align}
    \text{Si}(z)& =\text{si}(1,1,0) - \text{si}(1,1,z)=\int_0^z \frac{\sin(x)}{x}dx \,,\\
    \text{Ci}(z) &=-\text{ci}(1,1,z)= -\int_z^\infty\frac{\cos(x)}{x} dx \,.
\end{align}

With these ingredients at hand, we can write the final result of the superhorizon integrals as
\begin{align}
    \mathcal{I}_J^{\tilde{x}\ll 1}(u,v) & = C^\mathbb{1}  + \sum_{\vartheta \in \Omega_\vartheta} \left( C^s_{\vartheta} \sin(c_s \vartheta \xi_J) + C^c_{\vartheta} \cos(c_s \vartheta \xi_J) + C^{Si}_{\vartheta} \text{Si}(c_s \vartheta \xi_J) \right) \,,
    \label{eq:KernelIJ_super}
\end{align}
and
\begin{align}
    \mathcal{I}_Y^{\tilde{x}\ll 1}(u,v) = \mathcal{C}^\mathbb{1}  + \sum_{\vartheta \in \Omega_\vartheta} \Big( &\mathcal{C}^s_{\vartheta} \sin(c_s \vartheta \xi_Y) + \mathcal{C}^c_{\vartheta} \cos(c_s \vartheta \xi_Y) + \mathcal{C}^{Si}_{\vartheta} \text{Si}(c_s \vartheta \xi_Y) + \mathcal{C}^{\mathfrak{ci}}_{\vartheta} \mathfrak{ci}(1+2b,c_s \vartheta, \xi_Y) \Big) \,,
    \label{eq:KernelIY_super}
\end{align}
where the sum runs over $\Omega_\vartheta = \{u,v,u+v,u-v\}$, and the lenghty expressions for the coefficients ${C^i=C^i(u,v,b)}$ and $\mathcal{C}^i$ are provided explicitly in \cref{app:coefficients_super}.

\subsubsection{Subhorizon integrals}
In order to compute the large-$\tilde{x}$ contributions to the kernel, we proceed similarly to before by first using trigonometric identities to reduce products of sines and cosines. Due to the additional sine and cosine terms appearing in the subhorizon ($\tilde{x}\gg1$) expansion of the Bessel functions, cf.~\cref{eq:BesselJ_Largex,eq:BesselY_Largex}, the integrands now acquire a phase shift, as well as a larger number of terms than before, but are still given by powers of $\tilde{x}$ times sine or cosine.
The schematic form of the integrals is given in \cref{eq:KernelIJ_sub_integral}.
We define the generalized sine and cosine integrals, now including a phase shift $\Delta=b\pi/2$, as
\begin{align}
    \text{si}_\Delta(n,m,z,\Delta) \coloneq & \int_z^\infty dx \, x^{-n} \sin(m x + \Delta) \nonumber \\
    = & \cos(\Delta) \text{si}(n,m,z) + \sin(\Delta) \text{ci}(n,m,z) \,, \label{eq:si_D(nmzD)} \\
    \text{ci}_\Delta(n,m,z,\Delta) \coloneq & \int_z^\infty dx \, x^{-n} \cos(m x + \Delta) \nonumber \\
    = & \cos(\Delta) \text{ci}(n,m,z) - \sin(\Delta) \text{si}(n,m,z) \,, \label{eq:ci_D(nmzD)}
\end{align}
which recovers our previously defined expressions \cref{eq:si(nmz),eq:ci(nmz)} in the limit $\Delta \rightarrow 0$.
Using these definitions, we can express the result for the subhorizon integrals as follows
\begin{align}
    \mathcal{I}_J^{\tilde{x}\gg 1}(u,v) = \sum_{\tilde{\vartheta} \in \Omega_{\tilde{\vartheta}}} \Bigg(& \tilde{C}^s_{\tilde{\vartheta}} \sin((c_s \tilde{\vartheta}-1) \xi_J+\frac{b\pi}{2}) + \tilde{C}^c_{\tilde{\vartheta}} \cos((c_s \tilde{\vartheta}-1) \xi_J+\frac{b\pi}{2}) \nonumber \\
    &+ \tilde{C}^{si}_{\tilde{\vartheta}} \text{si}_\Delta\left(1+b,c_s \tilde{\vartheta} - 1,\xi_J,\frac{b \pi}{2}\right) \Bigg) \,,
    \label{eq:KernelIJ_sub}
\end{align}
and
\begin{align}
    \mathcal{I}_Y^{\tilde{x}\gg 1}(u,v) = \sum_{\tilde{\vartheta} \in \Omega_{\tilde{\vartheta}}} \Bigg(& \tilde{\mathcal{C}}^s_{\tilde{\vartheta}} \sin((c_s \tilde{\vartheta}-1) \xi_Y+\frac{b\pi}{2}) + \tilde{\mathcal{C}}^c_{\tilde{\vartheta}} \cos((c_s \tilde{\vartheta}-1) \xi_Y+\frac{b\pi}{2}) \nonumber \\
    &+ \tilde{\mathcal{C}}^{ci}_{\tilde{\vartheta}} \text{ci}_\Delta\left(1+b,c_s \tilde{\vartheta} - 1,\xi_Y,\frac{b \pi}{2}\right) \Bigg) \,,
    \label{eq:KernelIY_sub}
\end{align}
where $\Omega_{\tilde{\vartheta}} = \{0,u,v,-u,-v,u+v,u-v,-u+v,-u-v\}$.
We present the explicit coefficients $\tilde{C}^i$ and $\tilde{\mathcal{C}}^i$ in \cref{app:coefficients_sub}.

In summary, with \cref{eq:KernelIJ_super,eq:KernelIY_super,eq:KernelIJ_sub,eq:KernelIY_sub} we have obtained analytical expressions for all the contributions to the kernel $I(x,k,u,v)$ as defined in \cref{eq:KernelI_split,eq:KernelI_sub_super_split}.
At this point, let us emphasize the advantages of obtaining an analytical expression for the kernel, despite the lengthy coefficients. Firstly, this allows us to evaluate the spectrum fully analytically for a Dirac delta isocurvature spectrum, 
without having to implement costly numerical solvers. Secondly, it allows us to evaluate the kernel for arbitrary $b$,\footnote{Note, that the exact cases $b=\pm 1/2$ require some additional care, as the expansion of the Bessel function ${Y_{b+\frac{1}{2}}(x\ll1)}$ in \cref{eq:BesselY_Smallx} for general $b$ is ill-defined there and instead a logarithm arises, see \cref{eq:BesselY_Smallx_b_0pt5}. Also, the case $b=0$ needs to be treated separately, because logarithms appear in the kernel in this case as well \cite{Domenech:2021and}.} without having to recompute any numerical integrals on a case-by-case basis, and further gives us an analytical handle on some properties of the resulting induced GW spectrum, like the infrared (IR) scaling discussed below. Finally, for more complicated isocurvature power spectra, the induced GW spectrum has to be computed numerically, in which case it can also be favorable to have an analytical integrand in order to avoid building up numerical errors by nested numerical integrations.
We can now discuss the computation of the spectral GW energy density from our newly found kernel.

\section{Induced GW spectrum} \label{sec:Induced_GW_spectrum}
Let us now evaluate the induced GW spectrum.
Before that, however, note that we have not fixed the expansion history of the Universe after the $w$-dominated era, which depends on whether it transitions directly to the radiation-dominated era or goes through an intermediate eMD phase.
To be as general as possible, we proceed as follows. We assume an instantaneous transition for the $w$-dominated universe to a radiation-dominated universe at $\tau=\tau_{c}$ before any matter-dominated phase is reached,\footnote{Note that even though the matter-$w$ equality is never reached, we can still use the time of the would-be matter-$w$ equality $x_{w \rm eq}$ as a pivot scale as it is set by the initial ratio of energy densities.} that is $x_{w \rm eq}\geq x_c \gg 1$. We then match the solution of the induced GWs in the $w$-dominated era to the free tensor modes in the radiation-dominated era, as in \cite{Domenech:2020kqm}. Strictly speaking, one should match the kernels, i.e., match the curvature perturbation $\Phi$ and the tensor modes' Green’s function as in \cite{Altavista:2023zhw}. They are, however, equivalent for $x_c\gg 1$, up to oscillatory artifacts in the latter due to the sharp transition in the source term. Such oscillatory artifacts disappear once the transition becomes smoother \cite{Altavista:2023zhw}. In the end, one can take the oscillation average of the kernel at the end of the $w$-dominated era to evaluate the spectral density $\Omega_{\rm GW}(k)$ at the onset of the radiation-dominated phase \cite{Domenech:2020kqm}.

This approximation is valid if there is no intermediate eMD phase, which would be the case if the matter component decays before dominating (e.g.~PBHs and Q-Balls evaporate) or if it is the standard cold dark matter, which dominates much later. In the presence of an intermediate eMD phase between $w$-domination and radiation domination, one should set $x_c=x_{w \rm eq}$ and include a redshift factor $a_{w \rm eq}/a_{\rm end}$ both in the spectral density and peak frequency, where $a_{\rm end}$ is the scale factor at the end of the eMD phase. For example, if the early matter era is driven by evaporating ultra-light PBHs, the redshift factor is fixed in terms of the PBH mass and initial abundance \cite{Domenech:2024wao}. Although the duration of the eMD phase affects the amplitude and peak position of the induced GW spectrum, it does not change the overall spectral shape.

With the above assumption, we compute the spectral GW energy density $\Omega_{\rm GW}(k)$ by taking the oscillation average of the kernel $I(x,k,u,v)$ at a time $x_c=k\tau_c$, with $x_{w \rm eq}\geq x_c \gg 1$, when all tensor modes of interest are deep inside the horizon and behave as GWs.
Then we can take the $x\gg 1$ limit of the Bessel functions in \cref{eq:BesselJ_Largex,eq:BesselY_Largex} to obtain the oscillating behavior. Under the assumption that the oscillation timescale is much shorter than the decay timescale (determined by the power-law in $x$), we can compute the oscillation average as an integral over half a period, $(x_i,x_i+\pi)$, divided by $\pi$. We find
\begin{equation}
    \overline{Y^2_{b+\frac{1}{2}}(x\gg 1)}\approx \frac{1}{\pi x} \approx \overline{J^2_{b+\frac{1}{2}}(x\gg 1)} \quad \text{and} \quad \overline{Y_{b+\frac{1}{2}}(x\gg 1) J_{b+\frac{1}{2}}(x\gg 1)}\approx 0 \,,
\end{equation}
and thus, we obtain that the oscillation-averaged kernel is given by
\begin{align}
    \overline{I^2(x_c,u,v)} \approx \frac{\kappa^{4 b-4} }{x_c^{2+2 b} \pi} \Bigg(\bigg(\mathcal{I}_J^{\tilde{x}\ll 1}(u,v)+ \mathcal{I}_J^{\tilde{x}\gg 1}(u,v)\bigg)^2 + \bigg(\mathcal{I}_Y^{\tilde{x}\ll 1}(u,v) + \mathcal{I}_Y^{\tilde{x}\gg 1}(u,v)\bigg)^2 \Bigg) \,,
     \label{eq:KernelI_osc_avg}
\end{align}
with the sub- and superhorizon pieces of $\mathcal{I}_J$ and $\mathcal{I}_Y$ given in \cref{eq:KernelIJ_super,eq:KernelIY_super,eq:KernelIJ_sub,eq:KernelIY_sub}, respectively.

The spectral GW energy density then reads \cite{Kohri:2018awv, Domenech:2021ztg}
\begin{align}
    \Omega_{\rm GW}(k)= \frac{2 x_c^2}{3(1+b)^2} \int_{0}^{\infty} dv \int_{|1-v|}^{1+v} du \left(\frac{\left(1+v^2-u^2\right)^2-4 v^2}{4 u v}\right)^2
    {\cal P}_{S}(u k)
    {\cal P}_{S}(v k)
    \overline{I^2(x_c,u,v)} \,.
    \label{eq:Omega_GW}
\end{align}
Note that the resulting overall factor $x_c^{-2b} = (k\tau_c)^{-2b}$ after inserting \cref{eq:KernelI_osc_avg} in \cref{eq:Omega_GW} reflects the fact that the GW energy density redshifts faster (slower) than the $w$-background for a soft (stiff) equation of state.
Let us remind the reader that \cref{eq:Omega_GW} is valid for an instantaneous transition to a radiation-dominated universe right after the $w$-dominated phase. For an intermediate eMD era, one should add an additional factor $a_{w \rm eq}/a_{\rm end}$ due to the different redshift of the background.
In order to evaluate the remaining momentum integrals in $u$ and $v$ in \cref{eq:Omega_GW}, we need to specify the power spectrum $\mathcal{P}_S$ of the initial isocurvature perturbation $S_i$.

\subsection{GW spectrum for sharply peaked isocurvature spectrum}
For the sake of analytical simplicity, let us consider a Dirac delta isocurvature spectrum
\begin{equation}
    \mathcal{P}_S (k) = A_S \, \delta ( \ln (k/k_p) ) \,.
    \label{eq:Isocurvature_PS}
\end{equation}
Such a spectrum arises, for example, in the limit of a log-normal distribution with vanishing width, and may thus provide some physical insights also for more physically motivated, sharply peaked spectra. Such a sharp spectrum could plausibly be generated in the presence of narrow resonances during inflation \cite{Cai:2018tuh,Chen:2019zza}. Note that although Refs.~\cite{Cai:2018tuh,Chen:2019zza} discuss the curvature perturbation, it is straightforward to apply their formalism to spectator fields.

The Dirac delta parametrization of the power spectrum has the crucial advantage that one can trivially evaluate the momentum integrals over $u$ and $v$ in the GW spectrum by simply replacing $u=v=k_p/k$ in the kernel, and accounting for the additional factor $(k/k_p)^{-2}$ due to the logarithm in the argument of \labelcref{eq:Isocurvature_PS}. Respecting momentum conservation imposes a sharp cutoff of the spectrum above $k/k_p = 2$.

The induced GW spectrum is thus given by
\begin{align}
    \Omega_{\rm GW}(k/k_p) = & \frac{2A_S^2}{3\pi (1+b)^{2+2b}} \left(\frac{k_c}{k_{w \rm eq}}\right)^{2b} \left(\frac{k_p}{k_{w \rm eq}}\right)^{2 b-4} \nonumber\\& \times  \left(1-\frac{k^2}{4 k_p^2}\right)^2 \left(\frac{k}{k_p}\right)^{2 b-6}\left(\mathcal{I}_J^2(k_p/k) + \mathcal{I}_Y^2(k_p/k) \right)\Theta(k-2 k_p) \,,
    \label{eq:iGW_spectrum}
\end{align}
where we used that $k_c\simeq (1+b)/\tau_c$. Recall that $k_p\gg k_c\geq k_{w\rm eq}$.
The amplitude of the spectrum is mainly determined by the amplitude of the isocurvature power spectrum $A_S$, and the ratio of the peak scale to the one corresponding to equality, $k_p/k_{w \rm eq}$.
Current constraints on small scales still allow for relatively large values of $A_S$ \cite{Passaglia:2021jla,Domenech:2021and}.
In \cite{Domenech:2021and} it was found that in the radiation case the peak amplitude of the spectrum is suppressed by a factor $(k_p/k_{w \rm eq})^{-4}$ compared to the adiabatic case, and therefore a much larger amplitude of the initial isocurvature, $A_S$, is required to produce an observable signal. Here, we find that the suppression factor is modified to $(k_p/k_{w \rm eq})^{-4+2b}$ and thus becomes weaker as the equation of state becomes softer. In addition, the peak amplitude is enhanced for a softer equation of state, indicating that even for small amplitudes of the primordial isocurvature, the GW signal may become large enough to be observable. 

It is interesting to note that the dependence of the amplitude of the isocurvature-induced GW spectrum \eqref{eq:iGW_spectrum} on $b$ is opposite to that in the adiabatic case. In the adiabatic case, a stiff equation of state enhances the peak amplitude of the induced GW spectrum with respect to the radiation-dominated case; the main relevant effect is the relative redshift \cite{Domenech:2019quo,Domenech:2021ztg}. In the isocurvature case, it is the soft equation of state that enhances the amplitude with respect to the radiation-dominated case; the relative redshift competes with the maximum transfer of isocurvature to the curvature perturbation. The latter turns out to be the dominant factor.
%
\begin{figure}
\centering
\subfloat[Soft EoS ($b\geq0$)]{\includegraphics[width=0.49\textwidth]{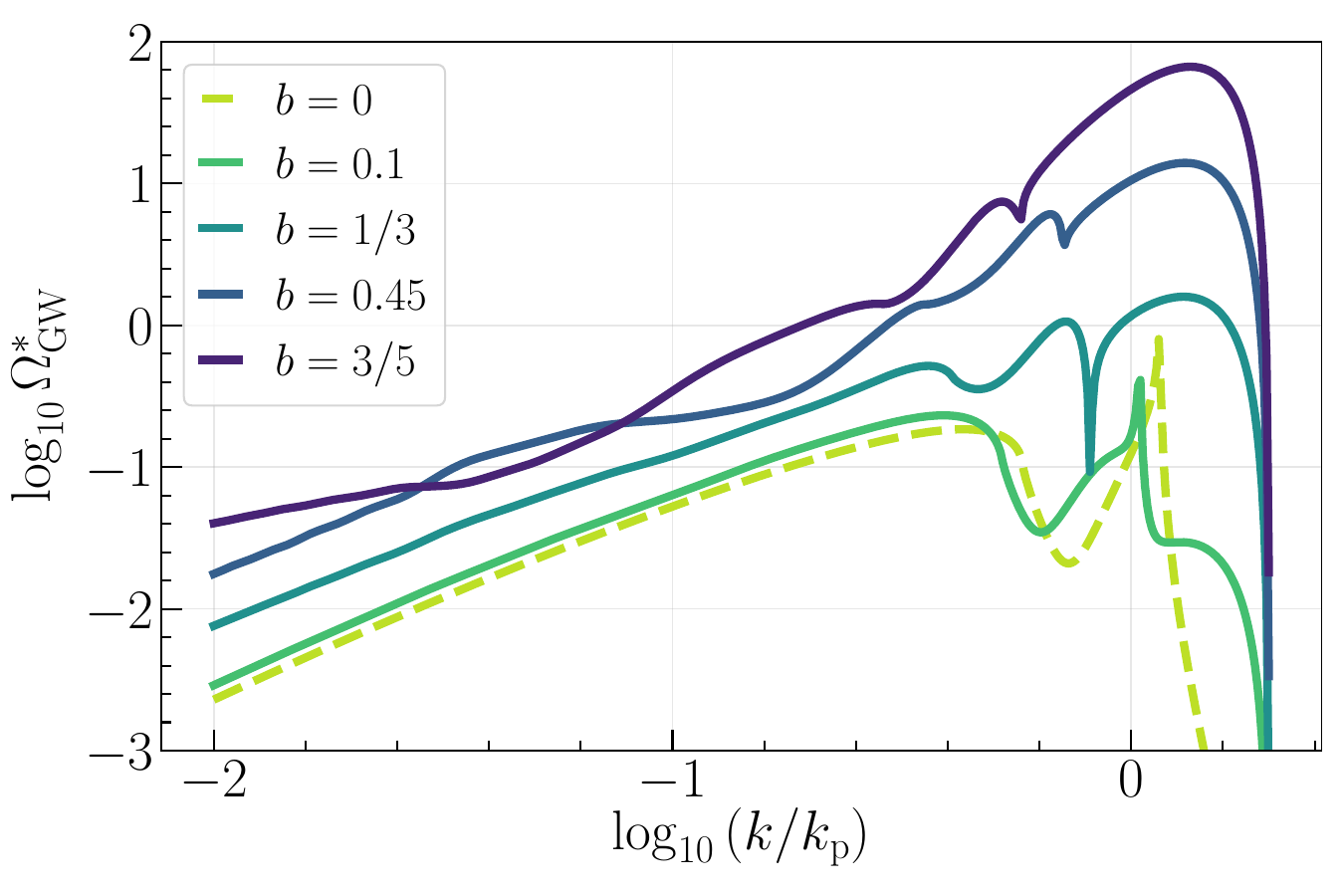}
\label{fig:subfig_Omega_GW_soft}}
\subfloat[Stiff EoS ($b\leq0$)]{\includegraphics[width=0.49\textwidth]{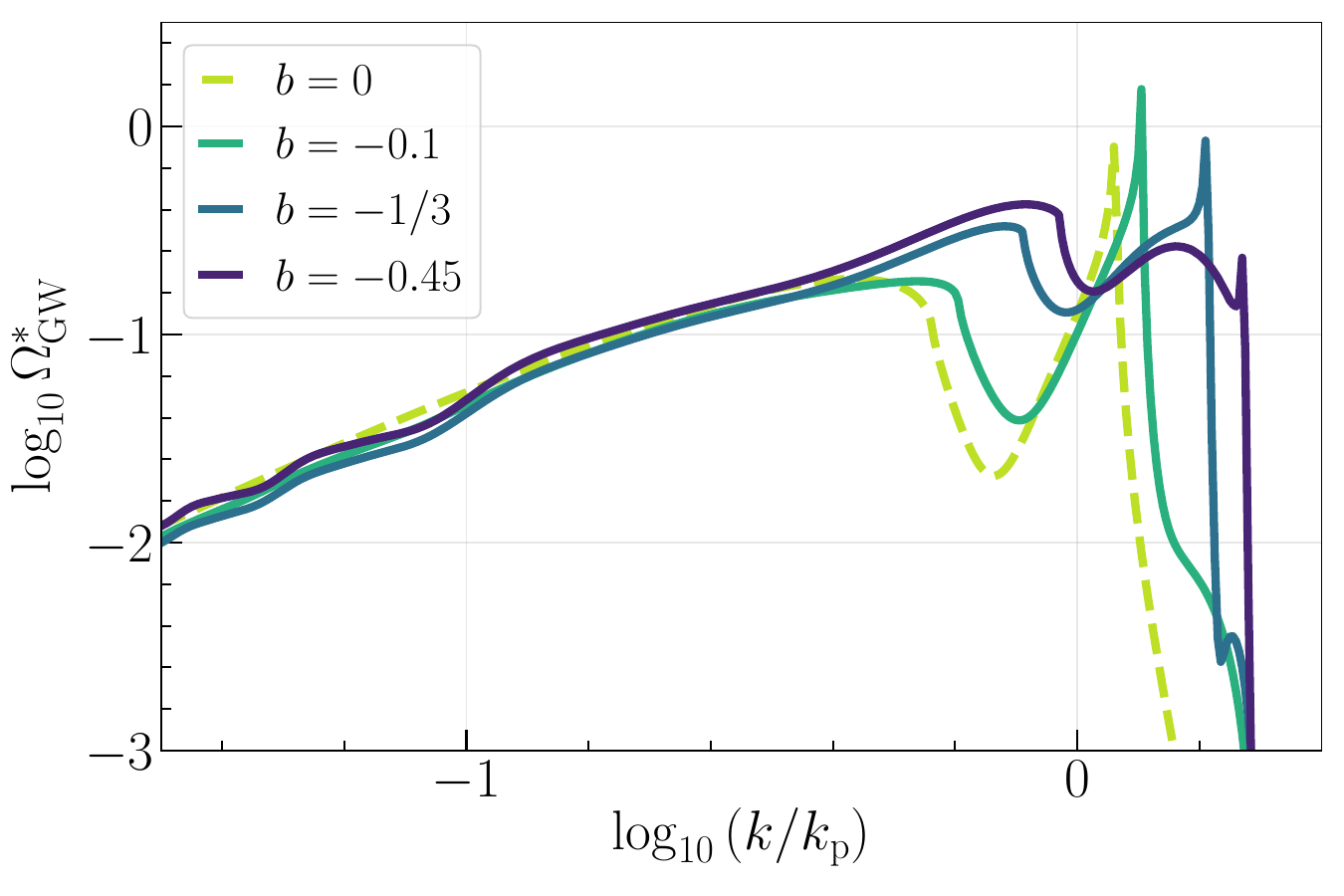}
\label{fig:subfig_Omega_GW_stiff}}
\caption{Here we show the effect of varying the equation of state $w$, encoded by the parameter $b$ defined in \cref{eq:bDefinition}, on the isocurvature-induced GW spectrum. Note how the peak amplitude is enhanced for soft EoS ($b>0$), and how the resonant scale located at $k/k_p = 2 c_s$ shifts with the varying speed of sound, $c_s=\sqrt{w}$. We show the normalized spectrum $\Omega_{\rm GW}^*$, cf.~\cref{eq:Omega_GW_norm}.}
\label{fig:OmegaGWPlot}
\end{figure}
%

In \cref{fig:OmegaGWPlot}, we display the normalized induced GW spectrum
\begin{equation}
    \Omega_{\rm GW}^*(k/k_p) \coloneq \frac{\Omega_{\rm GW}(k/k_p)}{A_S^2 \left({k_{w \rm eq}}/{k_c}\right)^{-2b} \left({k_p}/{k_{w \rm eq}}\right)^{2 b-4}} \,,
    \label{eq:Omega_GW_norm}
\end{equation}
as a function of the frequency relative to the peak frequency, $k/k_p$, showing in particular the effect of varying the EoS parameter $b$. Note that our analytical results use the approximation of splitting the induced GW generation between the (tensor) superhorizon and subhorizon regimes. Thus, there could be minor wiggles in the spectrum due to the sharp matching between those regimes, which we expect to be smeared out in the exact case. These effects are not relevant to understanding the main features of the isocurvature-induced GW spectrum. We start with a description of the spectral features and later, in \cref{sec:contributions_GW}, we discuss their physical origin.

First, for $b\approx 0$ we find that our result agrees excellently with the formulas derived in \cite{Domenech:2021and}, where early radiation domination ($b=0$) was assumed throughout, confirming the validity of the approximations used in our analytical computation. Furthermore, we observe a sharp feature in the spectrum located at $k/k_p = 2 c_s=2 \sqrt{w}$. For a stiffer (softer) EoS, this feature is shifted to higher (lower) frequencies, as the sound speed $c_s$ becomes larger (smaller). For a stiff EoS, it corresponds to a resonant peak, whereas for a soft EoS, it manifests itself instead as a destructive interference, and the peak of the spectrum remains near $k_p$. The resulting dip stems from a cancellation between the superhorizon and subhorizon source contributions, as we show later in \cref{sec:contributions_GW}.
Between $k/k_p = c_s$ and $k/k_p = 2 c_s$, the spectrum features a broad dip, which is most prominent for $b\lesssim 0$ and vanishes for $b\gtrsim 0.4$.

The scaling of the spectrum at small frequencies, $k\ll k_p$, can be understood as follows. By plotting separately the sub- and superhorizon contributions from $\mathcal{I}_{J/Y}(k_p/k)$, see \cref{fig:OmegaGW_sub_super}, we conclude that the IR behaviour is determined by the superhorizon contributions $\mathcal{I}_{J/Y}^{\tilde{x}\ll 1}$.
As shown in \cref{app:scaling_I_JY_superhorizon}, the dominant scaling in the IR is given by $\mathcal{I}_{J}^{\tilde{x}\ll 1}\sim \tilde{k}^4$ and $\mathcal{I}_{Y}^{\tilde{x}\ll 1}\sim \tilde{k}^4 + \tilde{k}^{4-2b}$, where we defined $\tilde{k}\coloneq k/k_p$.
This behavior can also be understood directly from the expressions \cref{eq:KernelIJ_super,eq:KernelIY_super}. Looking for the coefficients with the highest powers in $u = v = \tilde{k}^{-1} \gg 1$, we find that the terms $C^\mathbb{1}$, $\mathcal{C}^\mathbb{1}$, as well as $\mathcal{C}^{\mathfrak{ci}}_\vartheta$ are the most relevant ones, all scaling as $\tilde{k}^{4}$, cf.~\cref{eq:coeff_C1,eq:coeff_calC1,eq:coeff_u_ci,eq:coeff_u+v_ci}. By expanding the hypergeometric function in $\mathfrak{ci}(n,m,z)$ for large $m$, we find that the $\mathfrak{ci}$-term contributes another factor $\tilde{k}^{-2b}$.
Taking these considerations into account, we find that the low-frequency tail of the (normalized) spectrum approximately scales as
\begin{equation}\label{eq:IRscalingomega}
    \Omega_{\rm GW}^*(\tilde{k}\ll 1) \approx A_{\rm IR}(b) \times \tilde{k}^{2+2b} \left(\frac{\tilde{k}^{-2b}-1}{2b}\right)^2 \sim
    \begin{cases}
        \tilde k^{2-2b} & \quad b>0\\
        \tilde k^{2} \ln^2 (\tilde{k}) & \quad b=0\\
        \tilde k^{2+2b} & \quad b<0
    \end{cases} \, .
\end{equation}
The prefactor $A_{\rm IR}(b)$ is $\mathcal{O}(1)$ in our range of interest, $-1/2\leq b\lesssim 0.65$. As a rough order of magnitude estimate, it can be approximated by the numerical fit
\begin{equation}
    A_{\rm IR}(b) \approx \left(\frac{1.63}{2.92 - b} + \frac{0.64}{(1.01 - b)^2}\right)  (1+b)^{-2b}\,.
\end{equation}
In the limit of $b=0$, we recover the logarithmic scaling found in \cite{Domenech:2021and}.
From \cref{eq:IRscalingomega}, we conclude that the IR scaling is equal to the adiabatic case \cite{Domenech:2020kqm,Cai:2019cdl}. This is to be expected as the universal IR scaling derived in Ref.~\cite{Cai:2019cdl} does not assume any form of matter, only a finite lifetime for the source. Thus, the IR scaling of the induced GW spectrum is independent of whether one has adiabatic or isocurvature initial conditions.

Before moving on to studying the different contributions to the spectrum in more detail, it is interesting to stress that while for $b=0$ the isocurvature and curvature induced GWs have similar spectral shape for the Dirac delta primordial spectrum (see, e.g., fig.~3 of \cite{Domenech:2021and}), the difference is more pronounced for $b>0$ (compare \cref{fig:OmegaGWPlot} with fig.~1 of \cite{Domenech:2019quo}). For $b<0$, the features in the spectrum are similar to the $b=0$ case.

\subsection{Contributions to the GW spectrum} \label{sec:contributions_GW}
Let us discuss the subhorizon and superhorizon contributions, defined in \cref{eq:KernelI_sub_super_split}, to the induced GW spectrum, as well as the contributions stemming from the Einstein tensor and the energy-momentum tensor of matter, separately. Let us clarify that with the term ``superhorizon" here we refer to the $\tilde{x}=k\tilde{\tau}\ll 1$ contribution in \cref{eq:KernelI_sub_super_split}, which represents the generation of a tensor mode with momentum $k$ while on superhorizon scales, by scalar modes which may have entered the horizon already, depending on whether $u,v\lesssim c_s^{-1}$ or $u,v\gg c_s^{-1}$, respectively. The ``subhorizon" term corresponds to the regime where $\tilde{x}=k\tilde{\tau}\gg 1$, that is, a tensor mode with momentum $k$ generated while on subhorizon scales.
%
\begin{figure}[ht]
\centering
\subfloat[$b\approx 0$]{
\includegraphics[width=0.49\textwidth]{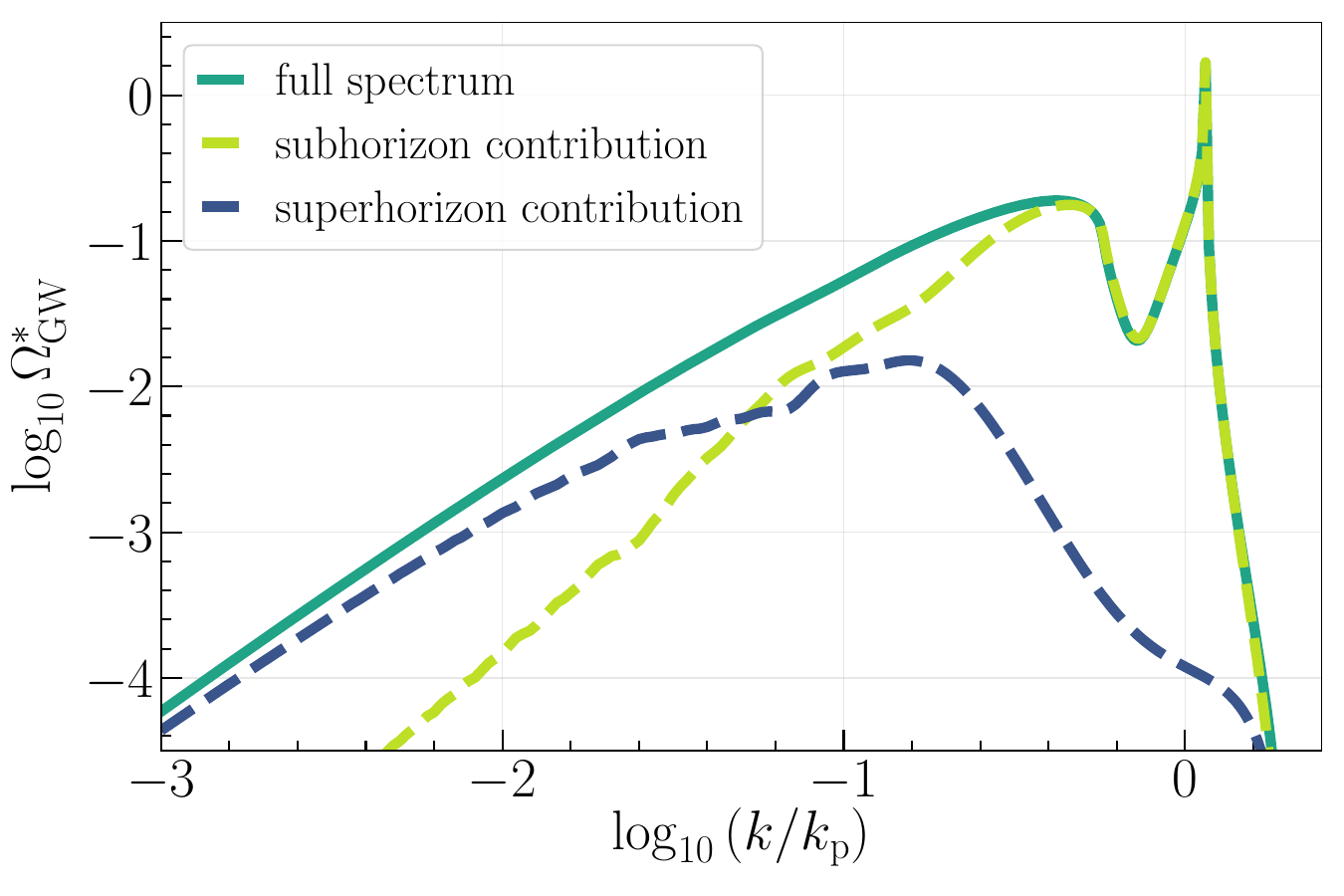}}
\subfloat[$b=-1/3$]{
\includegraphics[width=0.49\textwidth]{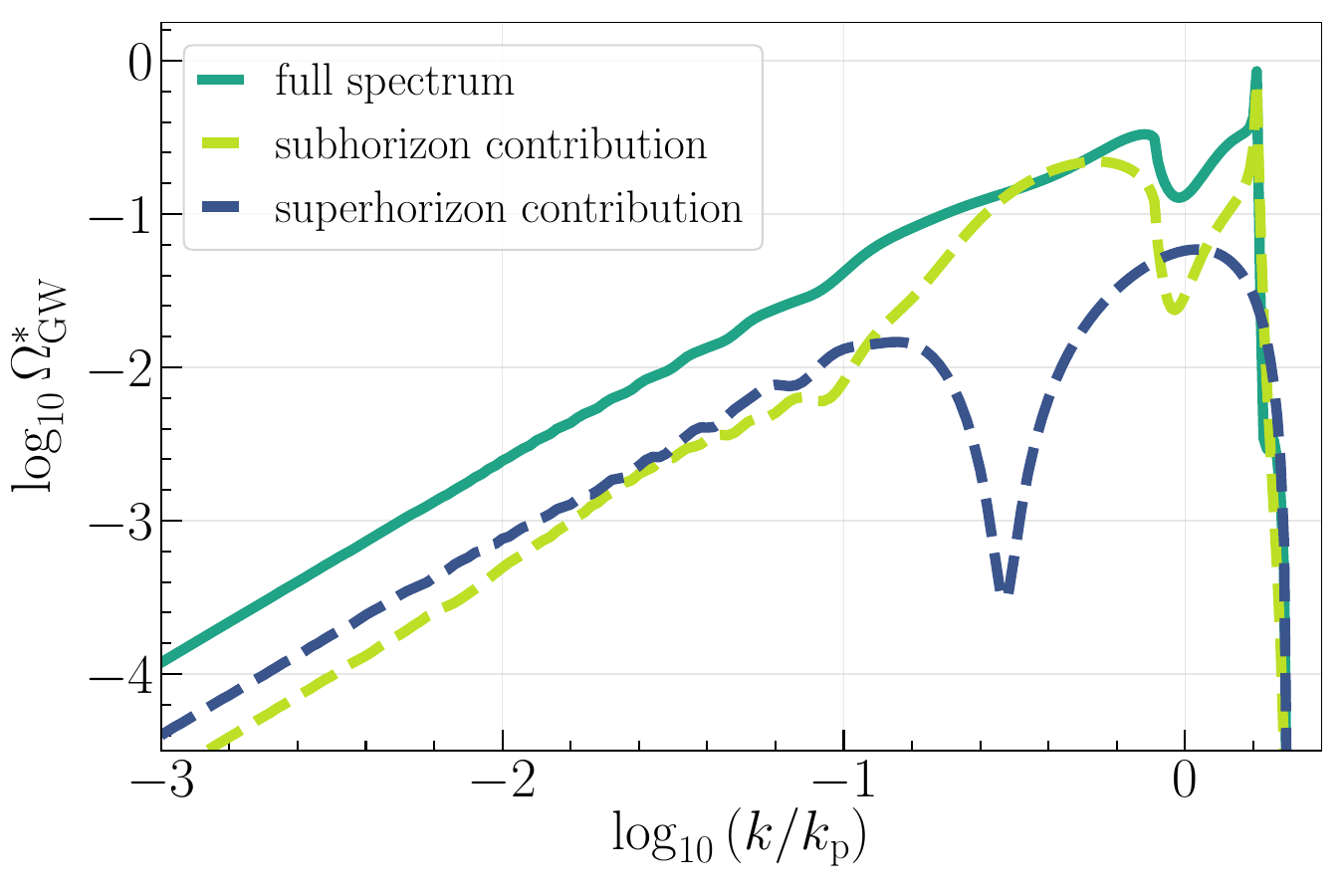}}
\qquad
\subfloat[$b=1/3$]{
\includegraphics[width=0.49\textwidth]{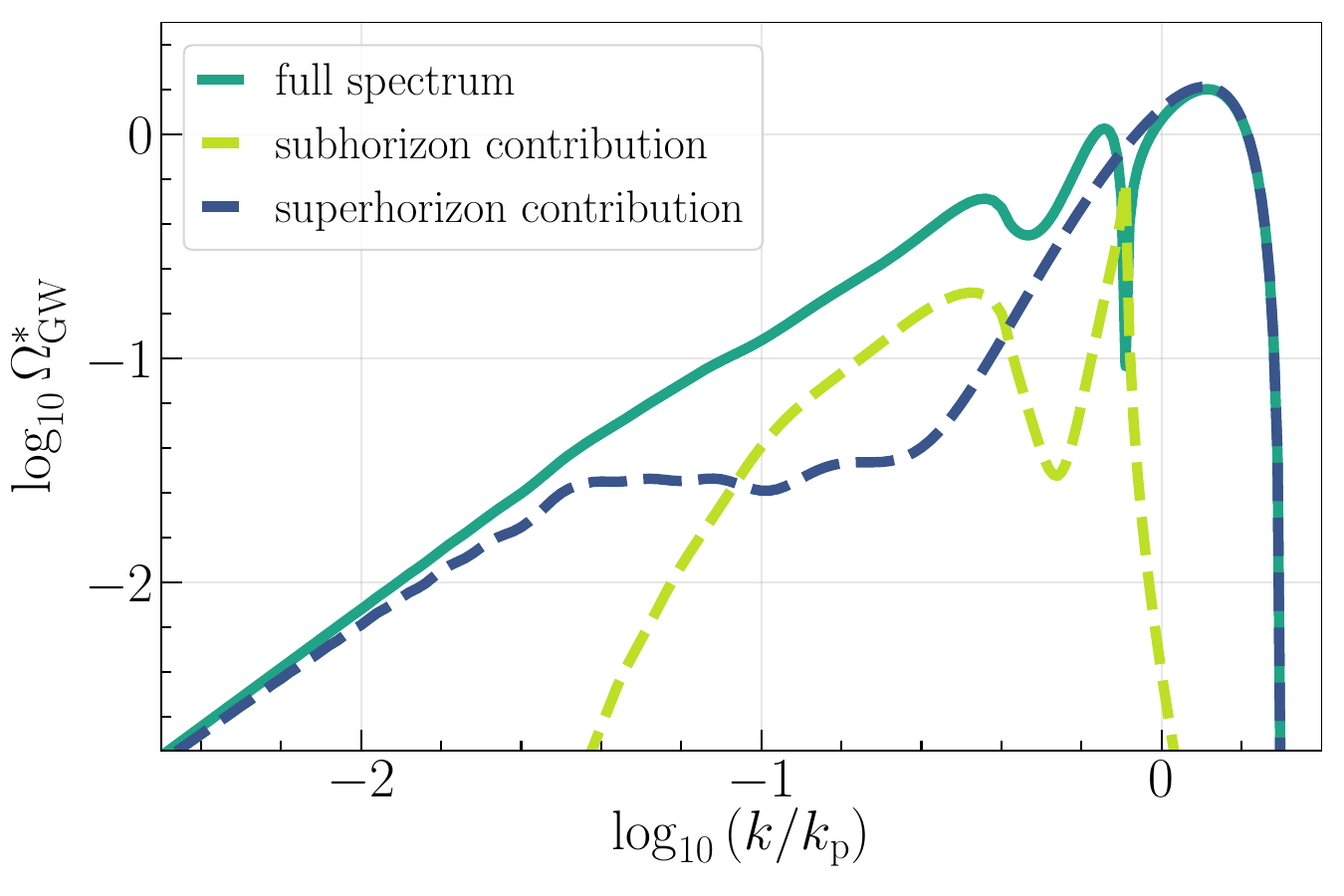}}
\subfloat[$b=3/5$]{
\includegraphics[width=0.49\textwidth]{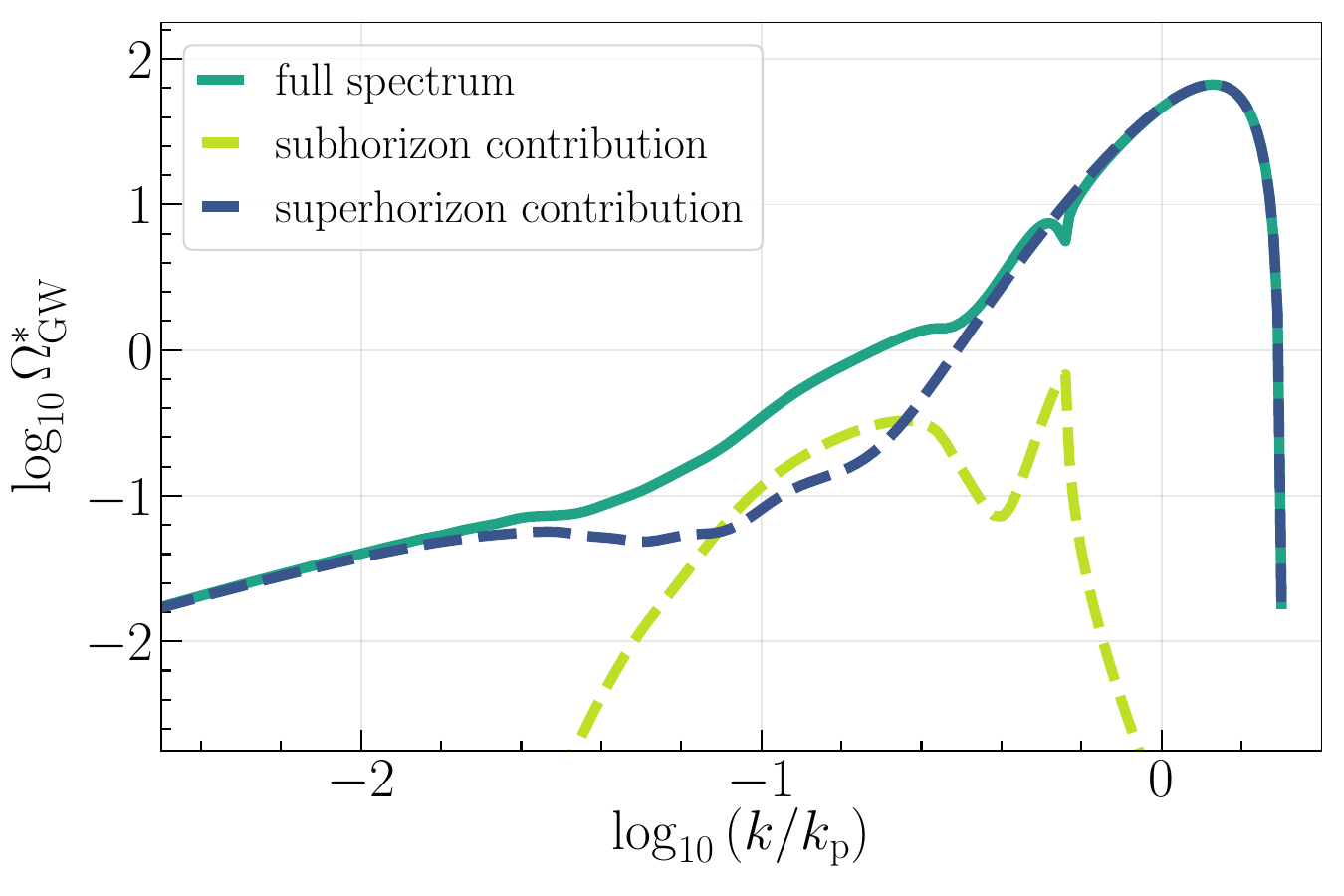}}
\caption{Comparison of the subhorizon and superhorizon contributions to the GW spectrum. We show the normalised spectrum $\Omega_{\rm GW}^*$. For numerical convenience, we evaluated the spectrum for $b\approx 0$ at $b=5\times 10^{-3}$.}
\label{fig:OmegaGW_sub_super}
\end{figure}
%
\begin{figure}
\centering
\subfloat[$b=1/3$]{
\includegraphics[width=0.49\textwidth]{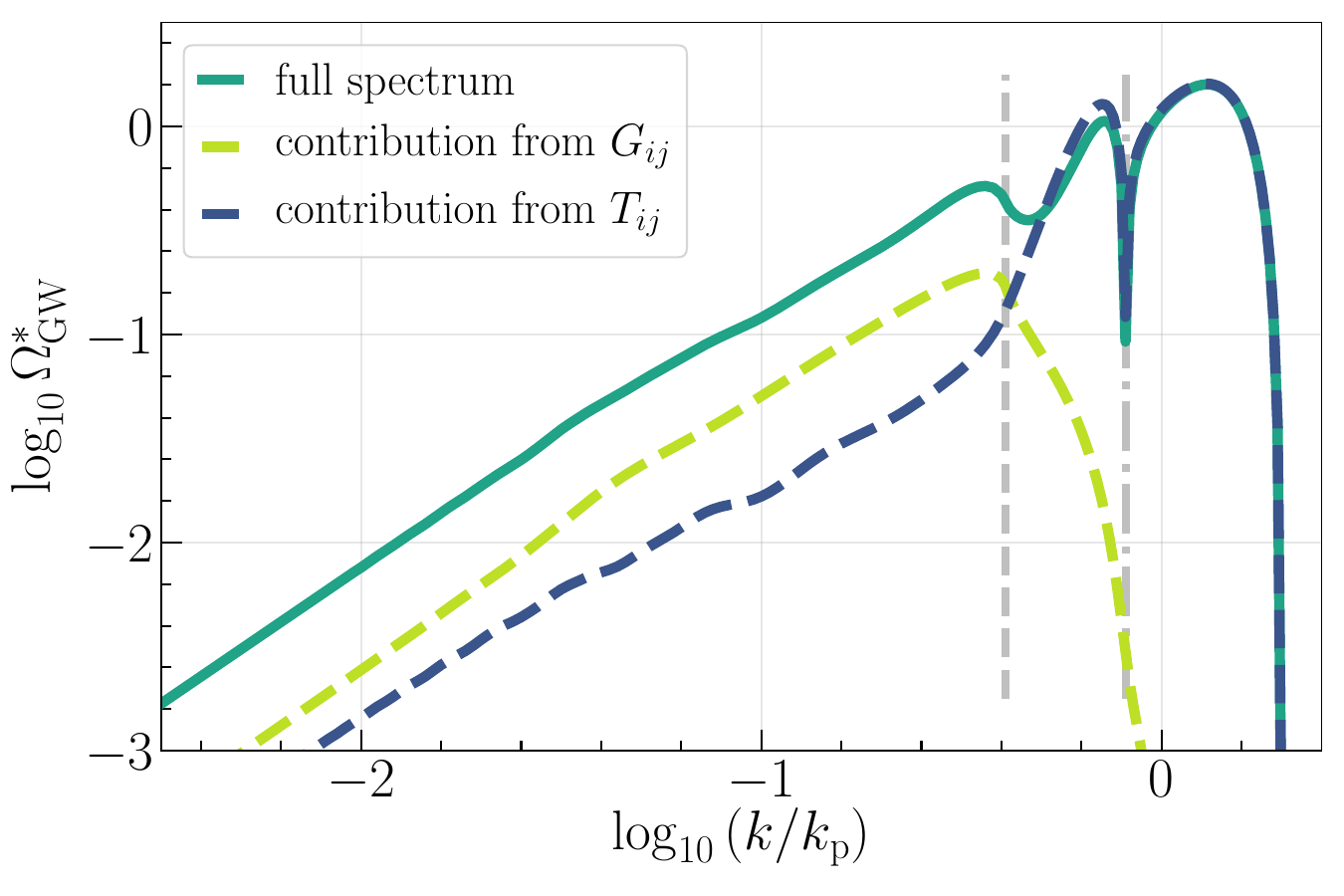}
\label{subfig:OmegaGW_G_vs_T_1by3}}
\subfloat[$b=-1/3$]{
\includegraphics[width=0.49\textwidth]{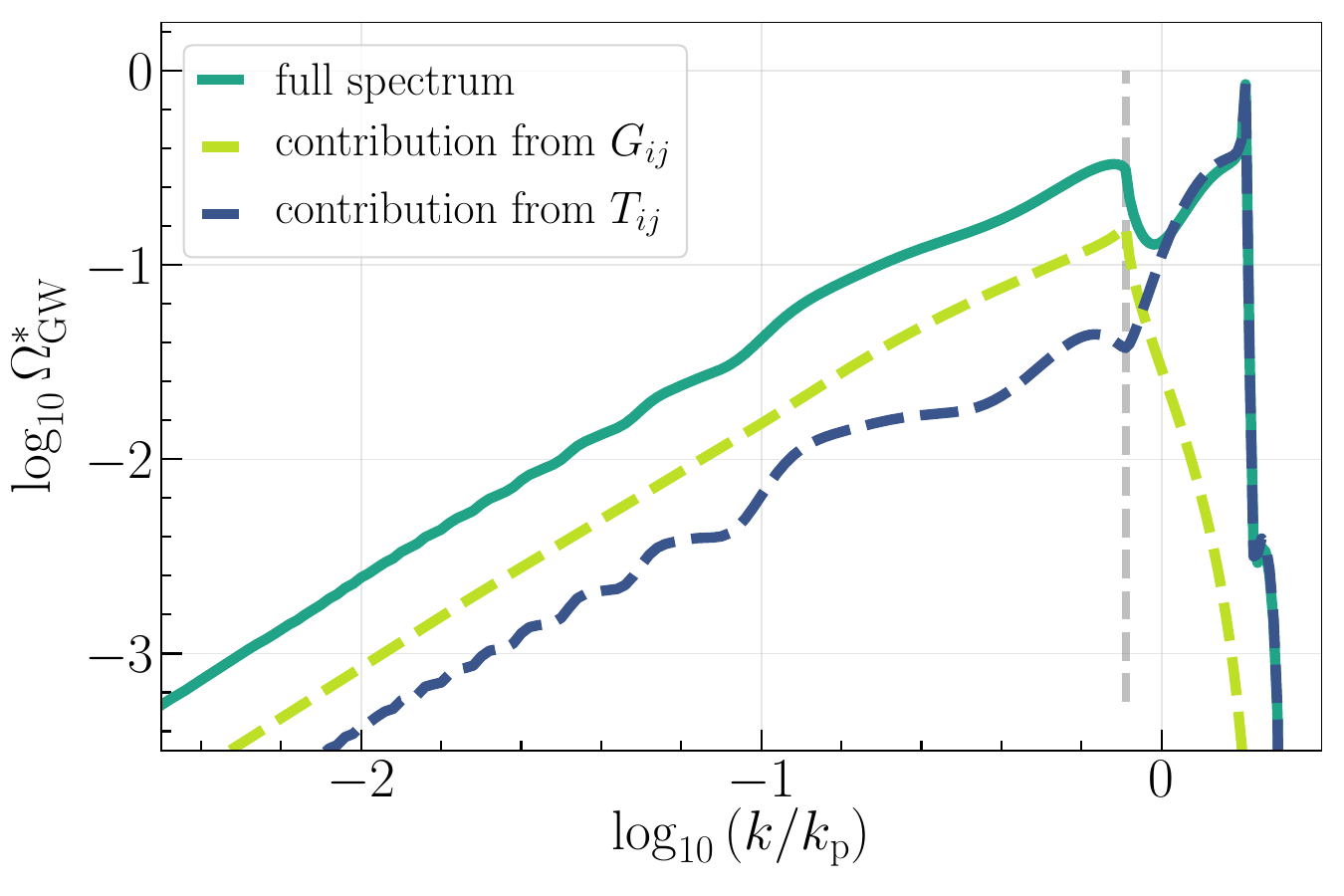}
\label{subfig:OmegaGW_G_vs_T_m1by3}}
\caption{Comparison of the contributions of the $\Phi\Phi$-term stemming from $G_{ij}$ and the $(\Phi+\Phi'/\calH)^2$-term stemming from $T_{ij}$. The vertical gray dashed line marks $k/k_p=c_s=\sqrt{w}$ for the respective values $w=1/6$ and $w=2/3$, and the dashdotted line in \cref{subfig:OmegaGW_G_vs_T_1by3} marks $k/k_p=2 c_s$.}
\label{fig:OmegaGW_G_vs_T}
\end{figure}
%

In \cref{fig:OmegaGW_sub_super} we compare the isolated contributions of the $\tilde{x}\ll 1$ (superhorizon) and $\tilde{x}\gg 1$ (subhorizon) expansions of the Bessel functions to the GW spectrum.
Interestingly, we find that for radiation and stiffer EoS (namely $b\leq0$), the peak regime is determined by the subhorizon contribution. It is also in the subhorizon part, where the resonance is effective.
On the other hand, for softer EoS, the superhorizon contribution becomes more important, and for large enough $b$, it fully dominates the spectrum in the peak region.
Independent of the EoS, the IR tail is determined by the superhorizon contribution.

To qualitatively understand the origin of this behavior, that is the relative size of the superhorizon contribution, recall that the isocurvature-induced curvature perturbation scales as ${\Phi_{\rm iso}(c_s x\gg 1)\propto (c_sx)^{-1-b}}$ and $\Phi_{\rm iso}(c_s x\ll 1)\propto (c_sx)^{1-b}$ in the large and small $x$ regimes, cf.~\cref{eq:Phi_iso_scaling}, respectively, and is largest around $c_sx\sim 1$. Thus, for stiff EoS with $b<0$, the slope is shallower for $c_sx\gg 1$ and the source is active longer in the subhorizon regime, while for $b>0$, the source is effective early on (i.e.~for $c_sx\ll1$), but decays more quickly after horizon entry. Large-scale, IR modes with $k\ll k_p$ enter the horizon long after the peak scale $k_p$ and are therefore mostly sourced at early times, that is, when $c_sx\ll 1$. However, for $b<0$, also the subhorizon curvature perturbation keeps sourcing superhorizon induced GWs (in our notation corresponding to $c_s x\gg 1$ and $\tilde x \ll 1$), as in the adiabatic case \cite{Domenech:2020kqm}.

The asymptotic scaling of the induced GW source in the superhorizon regime (i.e.~$\tilde x \ll 1$) also explains the large, broad peak near $k\sim k_p$ in the induced GW spectrum that appears for soft EoS. As it happens, for $b>0$ the conversion of isocurvature to curvature becomes more efficient compared to $b\leq 0$, in the sense that $\Phi$ approaches the constant adiabatic solution on super-sound-horizon scales, cf.~\cref{eq:Phi_ad_scaling}.
This leads to a bigger build-up of power on super-sound-horizon scales (corresponding to $c_svx\ll1$) with respect to the $b\leq 0$ case. The accumulation is naturally largest for $k\sim k_p$, i.e., for scales where the transfer from isocurvature to curvature is maximal. On sub-sound-horizon scales (that is, $c_svx\gg1$), the generation of tensor modes competes with the decay of the curvature perturbation. Due to the fast decay of $\Phi$ for $b>0$, the subhorizon resonance is not as efficient as for $b\leq0$. This is also the case for adiabatic initial conditions \cite{Domenech:2019quo,Domenech:2021ztg}. For $b \leq 0$, the superhorizon growth is completely buried under the more efficient resonant production on subhorizon scales. We provide a more quantitative discussion of the near-peak behavior for soft EoS in \cref{app:scaling_I_JY_superhorizon}. In particular, the approximate scaling of the spectrum near the peak is given in \cref{eq:Omega_GW_peak_soft}.

By looking at the different contributions in \cref{eq:KernelI_sub_super_split}, we can also understand the origin of the transition from the sharp peak feature at $k=2c_sk_p$ to a dip for the soft EoS case. The superhorizon contribution becomes larger for increasing $b>0$ and has an opposite sign compared to the subhorizon one. At around $b\approx 0.25$, both become comparable and a cancellation occurs. Thus, a dip appears. For larger $b$, the resonant peak of the subhorizon contribution becomes subdominant and, therefore, the dip becomes smaller.

In passing, it is also interesting to compare the contributions to the source term stemming from the Einstein tensor $G_{ij}$, with those coming from the energy-momentum tensor $T_{ij}$. In \cref{fig:OmegaGW_G_vs_T} we show the contributions from the $\Phi\Phi$-term (originating from $G_{ij}$), with that from $(\Phi+\Phi'/\calH)^2$ (proportional to the total velocity and stemming from $T_{ij}$). We find that the $T_{ij}$ contribution, i.e.~the term containing time derivatives, clearly dominates the peak regime for both soft and stiff EoS, whereas the term coming from $G_{ij}$ is larger in the IR tail regime.
The scale at which one becomes larger than the other is determined by the speed of sound and lies at $k/k_p=c_s=\sqrt{w}$. This also explains the prominent dip of the full spectrum between $k/k_p=c_s$ and $k/k_p=2 c_s$.

Physically, this can be interpreted such that at the smallest scales near $k_p$, sound waves and large velocity flows in the fluid are the dominant sources of GWs, while at large scales, the backreaction of large gradients of the gravitational potential (i.e.~scalar metric perturbations) due to the non-linear nature of general relativity is the dominant mechanism.

\section{Summary and conclusions}\label{sec:Summary}
In this paper, we have computed the induced GW spectrum from matter isocurvature initial conditions in a general cosmological background, applicable to standard cold dark matter as well as scenarios involving an eMD epoch.
To this end, we considered a two-fluid model with a dust-like matter component and a perfect fluid with a generic equation of state parameter $0<w\leq 1$, which could be realized in many different early universe scenarios involving, for example, a scalar field with a monomial or exponential potential, or the domination of discrete objects such as oscillons or PBHs, or superheavy metastable particles.
Observation of such a GW spectrum would thus offer a novel tool to probe the expansion history of the very early Universe and the reheating epoch between the end of inflation and the beginning of the standard radiation era.

In \cref{sec:Isocurvature} we discussed the generation and evolution of curvature perturbations sourced by initial isocurvature fluctuations. Importantly, we found a simple, yet very accurate, analytical expression for the gravitational potential, given in \cref{eq:Phi_Iso_GD}, which allowed us to analytically integrate the kernel of the induced GWs.
After showing the source term for the induced GWs in \cref{sec:Induced_GWs_Source}, we explained in some detail the computation of the kernel in \cref{sec:Induced_GWs_Kernel}. The key results are the analytical formulas for the super- and subhorizon contributions to $\mathcal{I}_{J/Y}$, given in \cref{eq:KernelIJ_super,eq:KernelIY_super,eq:KernelIJ_sub,eq:KernelIY_sub}, respectively, with the coefficients provided explicitly in \cref{app:coefficients}.

We then studied the induced GW spectrum in \cref{sec:Induced_GW_spectrum}. The result for a Dirac delta primordial isocurvature power spectrum is given in \cref{eq:iGW_spectrum}. Our expression allowed us to study the dependence on the EoS parameter $w$, shown in \cref{fig:OmegaGWPlot}. We find that $w$ determines the location of the resonant peak (or destructive interference dip) at $k=2c_s k_p$, a feature which could be crucial to determine the EoS parameter, if such a spectrum were observed. We note that although the location is the same as in the adiabatic case, the spectral shape of the isocurvature-induced GWs is different. In particular, the presence of a dip for $b>0$ instead of a resonant peak is a unique feature of the isocurvature-induced GWs. We further found that a soft EoS ($b>0$) gives rise to an enhanced GW peak amplitude compared to the radiation or stiff cases. 
Lastly, we showed that the low-frequency tail, given in \cref{eq:IRscalingomega}, follows the same scaling as for adiabatic initial conditions \cite{Domenech:2020kqm}. This agrees with the arguments for the universal infrared scaling of Ref.~\cite{Cai:2019cdl}.

An interesting application of our result for the induced GW kernel would be the generalization of the “universal” GWs from cosmological solitons described in \cite{Lozanov:2023aez,Lozanov:2023knf}. In scenarios where discrete objects such as solitons, oscillons, or PBHs come to dominate the early Universe, isocurvature perturbations are inevitably generated due to the shot noise arising from the random distribution of such objects in space, resulting in a characteristic $k^3$-isocurvature power spectrum \cite{Papanikolaou:2020qtd}. Therefore, in any such scenario, an isocurvature-induced GW spectrum, as the one we have computed, is expected to be generated, providing a valuable new observable to probe such models. From our work, we conclude that the low-frequency tail of the “universal” GWs from cosmological solitons will depend on the EoS. Namely, it will be modified by an additional factor of $k^{-2|b|}$. Most interesting would be to study the implications of the large, broad peak near $k\sim k_p$ that appears for $b>0$.
Unfortunately, for such a Poissonian power spectrum, the momentum integrals in \cref{eq:Omega_GW} likely cannot be computed analytically. We plan to implement the necessary numerical integrations for this case in a subsequent work.

Another related, interesting step will be to connect the model parameters (such as the isocurvature amplitude $A_s$ or the equality scale $k_{w \rm eq}$) to specific scenarios, such as PBH reheating or oscillon domination. In such a case, the peak amplitude and frequency will be related to the initial PBH or oscillon mass and abundance. Such relations will allow us to probe and constrain a wide class of models involving early matter domination.

\begin{acknowledgments}
    This research is supported by the DFG under the Emmy-Noether program grant no.~DO 2574/1-1, project number 496592360, and by the JSPS KAKENHI grant No.~JP24K00624.
\end{acknowledgments}

\appendix
\section{Background and perturbation equations \label{app:einsteinequations}}
The expansion of the homogeneous and isotropic background spacetime is governed by the Friedmann equations
\begin{align}
    3 \calH^2 \Mpl^2 & = a^2(\rho_{\rm m}+\rho_{w}) \,, 
    \label{eq:FriedmannEq} \\
    \left( 2 \calH' + \calH^2 \right) \Mpl^2 & = -a^2 w \rho_w  \,,
    \label{eq:FriedmannEq2}
\end{align}
where $a$ is the scale factor of the FLRW metric \labelcref{eq:FLRW}, $\calH$ is the conformal Hubble parameter $\calH = a' / a$, and $\rho_{\rm m}$ and $\rho_w$ denote the energy densities of the matter and primordial fluids, respectively.
Deep in the $w$-dominated era, i.e.~for $a\ll a_{w \rm eq}$, $\calH$ is given by $\calH = (1+b)/\tau$ in terms of conformal time $\tau$.
In this regime, the time dependence of the scale factor $a$ is given by
\begin{align}
   \frac{a}{a_{w \rm eq}} \approx \left(\sqrt{2}(1+b)\kappa\right)^{-1-b} x^{1+b} \,.
   \label{eq:ScaleFactor_wDE}
\end{align}

Turning to the first order perturbations, we write the perturbed fluid 4-velocities as $u_n^\mu=\bar{u}_n^\mu + \delta u _n^\mu$ with $(\bar{u}_n^\mu)=(1/a,\Vec{0})$ and $(\delta u_n^\mu)=(\Phi/a,\partial^i V_n / a)$  \cite{Malik:2008im}, where we have kept only the scalar part of the spatial velocity perturbations, and $n\in\lbrace \text{m},\, w \rbrace$.

Inserting the perturbed FLRW metric \cref{eq:FLRW} into the Einstein equations yields
\begin{align}
    \left(6 \calH \Phi' + 6 \calH^2 \Phi - 2 \Delta \Phi  \right) \Mpl^2 =& a^2 \left( \delta \rho_{\rm m} + \delta \rho_{w} \right) \,, \label{eq:00EinsteinEq} \\
    \left(\Phi' + \calH \Phi \right) \Mpl^2 =& \frac{1}{2} a^2 \left( V_{\rm m}\rho_{\rm m} + (1+w)V_{w} \rho_w \right) \,,  \label{eq:0iEinsteinEq} \\
    \left(\Phi'' + 3\calH \Phi' + (\calH^2 + 2 \calH')\Phi  \right) \Mpl^2 =& -\frac{1}{2} a^2 c_{w}^2 \delta \rho_w \,,
     \label{eq:ijEinsteinEq}
\end{align}
for the $00$, $0i$ and $ij$ trace components at linear order. Here, $c_w$ denotes the sound speed of the $w$-fluid.
For convenience, we will set $\Mpl=1$ from now on.

The energy conservation equations are
\begin{align}
    \delta \rho_{\rm m}' + 3 \calH \delta \rho_{\rm m} + \rho_{\rm m} ( 3 \Phi' + \Delta V_{\rm m} ) =& 0 \,, \label{eq:Energy_Cons_m}  \\
    \delta \rho_w ' + 3 (1+c_w^2)\calH \delta\rho_w + (1+w) \rho_w (3\Phi' + \Delta V_w) =& 0 \,, \label{eq:Energy_Cons_w}
\end{align}
resulting from the $\nu=0$ component of the covariant conservation of the energy-momentum tensors $\nabla_\mu T^{\mu\nu}_n=0$ of the matter and $w$-fluids, respectively.
The spatial components $\nu=i$ give the momentum conservation equations
\begin{align}
    V_{\rm m}' + \calH V_{\rm m} =& \Phi \,, \label{eq:Momentum_Cons_m} \\
    V_w' + (1-3w)\calH V_w + \frac{c_w^2}{1+w}\frac{\delta \rho_w}{\rho_w} =& \Phi \,. \label{eq:Momentum_Cons_w}
\end{align}
Combining \cref{eq:Energy_Cons_m} and \cref{eq:Energy_Cons_w}, one can derive an equation for the time evolution of the isocurvature perturbation $S$. Then, using also \cref{eq:Momentum_Cons_m,eq:Momentum_Cons_w}, one finds the second order equation
\begin{equation}
    S'' + (1+3(c_s^2-w))\calH S' - k^2(c_s^2-w)S = \frac{2 c_s^2}{a^2(1+w)\rho_w}k^4 \Phi\,,
    \label{eq:SEq}
\end{equation}
which, together with \cref{eq:PhiEq}, yields a closed system for the evolution of $\Phi$ and $S$.
The speed of sound in the two-fluid system is defined as
\begin{equation}
    c_s^2 = \, \frac{c_w^2 (1+w)\rho_w}{\rho_{\rm m}+(1+w)\rho_w} \,.
    \label{eq:sound_speed}
\end{equation}

%
\begin{figure}
\centering
\subfloat[Breakdown of the approximation near $\tau_{w \rm eq}$.]{
\includegraphics[width=0.49\textwidth]{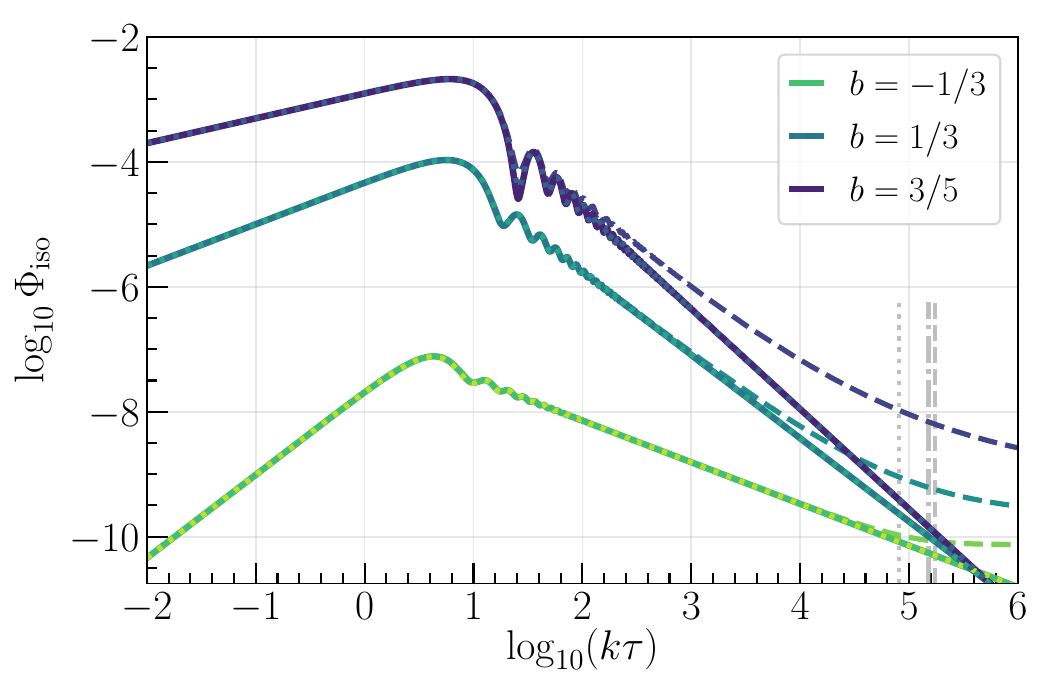} \label{subfig:Phi_1}}
\subfloat[Breakdown of the approximation for very soft EoS.]{
\includegraphics[width=0.49\textwidth]{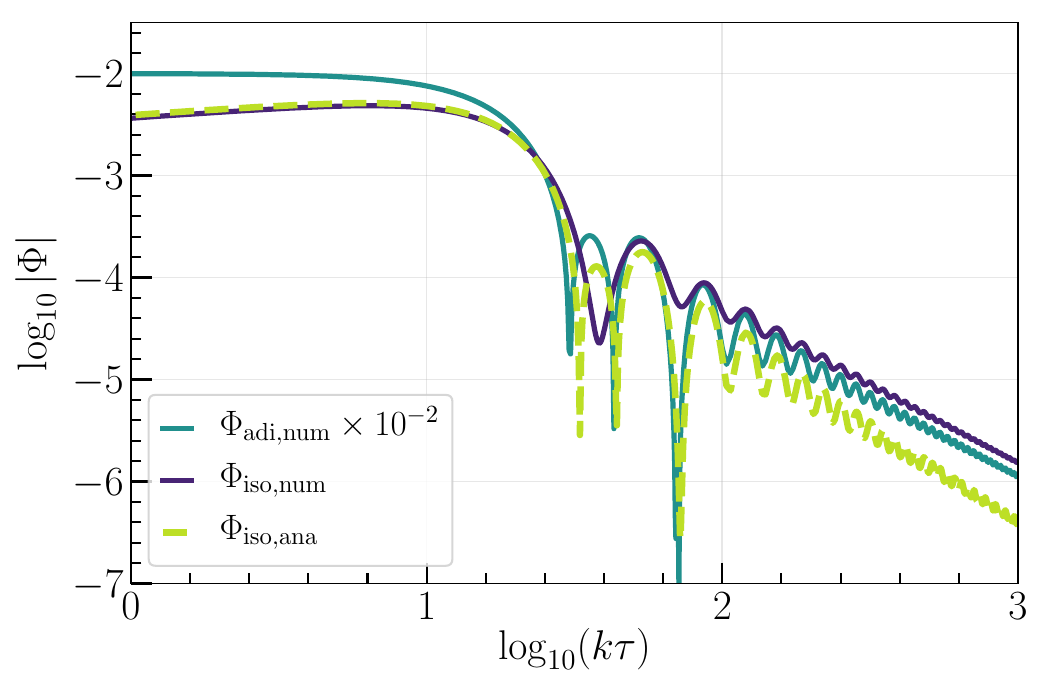} \label{subfig:Phi_2}}
\caption{\textit{\Cref{subfig:Phi_1}:} The evolution of the gravitational potential $\Phi_{\rm iso}(k\tau)$ for soft and stiff equation of state with $\kappa = 10^5$. Solid lines represent the approximate analytical solution \labelcref{eq:Phi_Iso_GD}, dotted lines show the exact analytical solution \cref{eq:Phi_iso_Green_sol}, and the dashed colored lines show a numerical solution of the full coupled system \cref{eq:PhiEq,eq:SEq}. The figure illustrates how the analytical approximation captures the exact numerical solution remarkably well for $\tau \ll \tau_{w \rm eq}$, with $k \tau_{w \rm eq}$ marked by the gray dotted, dash-dotted and dashed vertical lines for $b=-\frac{1}{3}, \, \frac{1}{3}$ and $\frac{3}{5}$, respectively. See how for soft EoS the approximation breaks down earlier, whereas for stiff EoS it is valid almost up to $\tau_{w \rm eq}$. \\
\textit{\Cref{subfig:Phi_2}:} Here we show $| \Phi_{\rm iso}(k\tau)|$ for a very soft EoS with $b=0.8$ and $\kappa = 10^8$. We compare the analytical approximation \cref{eq:Phi_Iso_GD} with numerical solutions for isocurvature and adiabatic initial conditions. See how the approximate solution tracks the isocurvature solution before horizon entry, but starts behaving similarly to the adiabatic one with zero-crossing oscillations for $x\gtrsim 1$ (note that we are plotting the absolute value).
}
\label{fig:Phi_iso_2}
\end{figure}
%

\section{Derivation of the induced GW kernel}\label{app:derivation_kernel}
In this appendix, we provide some more details on the computation of the kernel $I(x,k,u,v)$. First, the Green's function for tensor modes in a $w$-dominated background is obtained from the two homogeneous solutions of \cref{eq:Tensor_EoM}, and is given by \cite{Domenech:2021ztg}
\begin{equation}
    \mathcal{G}_h(x,\tilde{x}) = \frac{\pi}{2} \frac{\tilde{x}^{b+3/2}}{x^{b+1/2}} \left(Y_{b+\frac{1}{2}}(x) J_{b+\frac{1}{2}}(\tilde{x})-J_{b+\frac{1}{2}}(x) Y_{b+\frac{1}{2}}(\tilde{x})\right) \Theta(x-\tilde{x}) \, ,
    \label{eq:TensorGreensFctGen}
\end{equation}
with the Bessel functions of the first and second kind, $J_\nu$ and $Y_\nu$.

The source term $f(x, k, u, v)$ deep inside $w$-domination, which appears in \cref{eq:Kernel_I}, is given by
\begin{equation}
    f(x, k, u, v) = T_\Phi(v x)T_\Phi(u x) + \frac{1+b}{2+b}\left(T_\Phi(v x)+\frac{T_\Phi'(v x)}{\calH}\right) \left(T_\Phi(u x)+\frac{T_\Phi'(u x)}{\calH}\right)  \,,
    \label{eq:kernel_fxuv}
\end{equation}
resulting from \cref{eq:source_term_S_ab_wDE}. Here we defined the transfer function $T_\Phi$ by $\Phi_{\rm iso}= S_i \times T_\Phi$ with $\Phi_{\rm iso}$ given in \cref{eq:Phi_Iso_GD}.
Inserting the approximate transfer function for $\Phi_{\rm iso}$ from \cref{eq:Phi_Iso_GD} into \cref{eq:kernel_fxuv}, the explicit expression for the source term reads
\begin{align}
    f(x&,k,u,v) = \frac{9\ 2^{b-3} (b+1)^{2 b+2} w^2 }{(b+2) z_u^2 z_v^2}\kappa ^{2 b-2} x^{2-2 b}\Bigg((2 b+3)  \nonumber \\
    & \times \left(-\frac{15 b (b+3) j_2(z_u)}{4 \left(b^2+3 b-4\right)}+2 y_1(z_u)+\frac{2}{z_u^2}+1\right) \left(-\frac{15 b (b+3) j_2(z_v)}{4 \left(b^2+3 b-4\right)}+2 y_1(z_v)+\frac{2}{z_v^2}+1\right) \nonumber \\
    & +\left(\frac{15 (b+3) b j_2(z_u)}{4 (b+4)}+\frac{15 (b+3) b z_u j_3(z_u)}{4 (b-1) (b+4)}-\frac{2 (b+3)}{z_u^2}-2 b y_1(z_u)-2 z_u y_2(z_u) -(1+b) \right) \nonumber \\
    & \times \left(-\frac{15 b (b+3) j_2(z_v)}{4 \left(b^2+3 b-4\right)}+2 y_1(z_v)+\frac{2}{z_v^2}+1\right)+\left(-\frac{15 b (b+3) j_2(z_u)}{4 \left(b^2+3 b-4\right)}+2 y_1(z_u)+\frac{2}{z_u^2}+1\right) \nonumber \\
    &  \times \left(\frac{15 (b+3) b j_2(z_v)}{4 (b+4)}+\frac{15 (b+3) b z_v j_3(z_v)}{4 (b-1) (b+4)}-\frac{2 (b+3)}{z_v^2}-2 b y_1(z_v)-2 z_v y_2(z_v)-(1+b)\right) \nonumber \\
    &  +\frac{1}{b+1}\left(\frac{15 (b+3) b j_2(z_u)}{4 (b+4)}+\frac{15 (b+3) b z_u j_3(z_u)}{4 (b-1) (b+4)}-\frac{2 (b+3)}{z_u^2}-2 b y_1(z_u)-2 z_u y_2(z_u)-(1+b)\right)  \nonumber \\
    &  \times \left(\frac{15 (b+3) b j_2(z_v)}{4 (b+4)}+\frac{15 (b+3) b z_v j_3(z_v)}{4 (b-1) (b+4)}-\frac{2 (b+3)}{z_v^2}-2 b y_1(z_v)-2 z_v y_2(z_v)-(1+b)\right)\Bigg)\,,
    \label{eq:kernel_fxuv_expr}
\end{align}
where we defined $z_u = c_s u x$ and $z_v = c_s v x$ with $c_s =\sqrt{w}$.

The asymptotic expansions of the Bessel functions used in the split \cref{eq:KernelI_sub_super_split} are given by
\begin{align}
    J_{b+\frac{1}{2}}(x\ll 1) & \approx \frac{(x/2)^{b+\frac{1}{2}}}{\Gamma \left(b+\frac{3}{2}\right)}-\frac{(x/2)^{b+\frac{5}{2}}}{\left(b+\frac{3}{2}\right) \Gamma \left(b+\frac{3}{2}\right)} \,, \label{eq:BesselJ_Smallx}\\
    J_{b+\frac{1}{2}}(x\gg 1) & \approx \sqrt{\frac{2}{\pi }}\frac{\sin \left(x-\frac{\pi  b}{2}\right)}{\sqrt{x}} \,, \label{eq:BesselJ_Largex}
\end{align}
and
\begin{align}
    Y_{b+\frac{1}{2}}(x\ll 1) & \approx -\frac{(x/2)^{-b-\frac{1}{2}} \Gamma \left(b+\frac{1}{2}\right)}{\pi } + \frac{(x/2)^{b+\frac{1}{2}} \sin (\pi  b) \Gamma \left(-b-\frac{1}{2}\right)}{\pi }  \,, \label{eq:BesselY_Smallx}\\
    Y_{b+\frac{1}{2}}(x\gg 1) & \approx -\sqrt{\frac{2}{\pi }}\frac{ \cos \left(x-\frac{\pi  b}{2}\right)}{\sqrt{x}} \,. \label{eq:BesselY_Largex}
\end{align}
We keep only the leading order for the large-$x$ expansion, but include also the first subleading term for $x\ll 1$, as for $b\rightarrow -1/2$ the two terms in \cref{eq:BesselY_Smallx} become comparable, yielding a logarithm for $b=-1/2$.
Note also that for $b=1/2$ the second gamma function in \cref{eq:BesselY_Smallx} diverges. The correct expansions for $b=\pm 1/2$ are instead
\begin{align}
    Y_{0}(x\ll 1) & \approx \frac{2}{\pi}\left(\log(x/2) + \gamma_E\right)\,, \nonumber \\
    Y_{1}(x\ll 1) & \approx \frac{1}{2\pi x}\left(x^2 \left(\log (x^2/4)+2 \gamma_E -1\right)-4\right) \,,
    \label{eq:BesselY_Smallx_b_0pt5}
\end{align}
with the Euler-Mascheroni constant $\gamma_E\approx 0.577$.
We determine the matching points $\xi_J$ and $\xi_Y$ for the two kinds of Bessel functions by matching the leading power-laws of the large and small $x$ expansions of $J_{b+\frac{1}{2}}$ and $Y_{b+\frac{1}{2}}$, respectively. We find
\begin{equation}
    \xi_{J}= 2 \pi ^{-\frac{1}{2 (b+1)}} \Gamma \left(b+\frac{3}{2}\right)^{\frac{1}{b+1}} \quad \text{and} \quad \xi_Y= 2 \pi ^{-\frac{1}{2 b}} \Gamma \left(b+\frac{1}{2}\right)^{1/b} \,.
    \label{eq:matching_points}
\end{equation}
For our $b$-values of interest we have $\xi_J\gtrsim \xi_Y\sim \mathcal{O}(0.1-1)$.

\subsection{Superhorizon integrals}
In the computation of the integrals, we use trigonometric identities such as
\begin{align}
    \cos(\vartheta_1 \tilde{x}) \cos(\vartheta_2 \tilde{x}) & = \frac{1}{2} \left( \cos((\vartheta_1-\vartheta_2)\tilde{x}) + \cos((\vartheta_1 + \vartheta_2) \tilde{x})\right) \,, \nonumber \\
    \sin(\vartheta_1 \tilde{x}) \sin(\vartheta_2 \tilde{x}) & = \frac{1}{2} \left( \cos((\vartheta_1-\vartheta_2)\tilde{x}) - \cos((\vartheta_1 + \vartheta_2) \tilde{x})\right) \,, \nonumber \\
    \sin(\vartheta_1 \tilde{x}) \cos(\vartheta_2 \tilde{x}) & = \frac{1}{2} \left( \sin((\vartheta_1-\vartheta_2)\tilde{x}) + \sin((\vartheta_1 + \vartheta_2) \tilde{x})\right) \,, 
    \label{eq:trig_identities}
\end{align}
to reduce the number of sines and cosines in the expressions.
To simplify the equations we also used that $\tilde{x}^{-n} = \frac{d}{d\tilde{x}}\left(\frac{1}{1-n}\tilde{x}^{1-n}\right)$, and thus by partial integration
\begin{align}
    \int_{x_0}^{x_m} d\tilde{x} \, \tilde{x}^{-n} \sin(m \tilde{x}) & = \left. \frac{\tilde{x}^{1-n}}{1-n} \sin(m \tilde{x}) \right|_{x_0}^{x_m} - \int_{x_0}^{x_m} d\tilde{x} \, \frac{m \, \tilde{x}^{1-n}}{1-n} \cos(m \tilde{x}) \,.
    \label{eq:partial_int}
\end{align}
After using the trigonometric identities \cref{eq:trig_identities}, the integrals in the superhorizon term $\mathcal{I}_J^{\tilde{x}\ll 1}(u,v)$ can be brought to the form
\begin{align}
    \mathcal{I}_J^{\tilde{x}\ll 1}(u,v) & = \int_0^{\xi_J} d\tilde{x}\,  \left(\sum_{\lambda \in \Omega_\lambda^\mathbb{1} } c^\mathbb{1}_{\lambda} \tilde{x}^{-\lambda} +\sum_{\vartheta \in \Omega_\vartheta} \Big( \sum_{\lambda \in \Omega_\lambda^s } c^s_{\lambda,\vartheta} \tilde{x}^{-\lambda} \sin(c_s \vartheta \tilde{x}) + \sum_{\lambda \in \Omega_\lambda^c } c^c_{\lambda,\vartheta} \tilde{x}^{-\lambda} \cos(c_s \vartheta \tilde{x}) \Big) \right) \,,
    \label{eq:KernelIJ_super_integral}
\end{align}
where $\Omega_\vartheta = \{u,v,u+v,u-v\}$ and $\Omega_\lambda^\mathbb{1} = \{-2,0,2,4\}$, $\Omega_\lambda^s = \{-1,1,3,5\}$, $\Omega_\lambda^c = \{-2,0,2,4,6\}$, and the prefactors $c_{\lambda}^i$ are constants.
Note that although some of the integrals in \cref{eq:KernelIJ_super_integral} would diverge at the origin if considered individually, in the overall sum the divergent pieces exactly cancel, leaving the total result finite.

The integrand of $\mathcal{I}_Y^{\tilde{x}\ll 1}(u,v)$ can be written in the same way as \cref{eq:KernelIJ_super_integral}, namely as a sum of sine and cosine terms. However, due to the different power of $x$ in the leading order of the expansion of $Y_{b+\frac{1}{2}}(x\ll 1)$ in \labelcref{eq:BesselY_Smallx} compared to \labelcref{eq:BesselJ_Smallx}, the powers of $\tilde{x}$ in the integrand now depend on $b$.
In the integrals stemming from the leading order term we find $\Omega_\lambda^\mathbb{1} = \{2b+1,2b+3,2b+5\}$, $\Omega_\lambda^s = \{2b+2,2b+4,2b+6\}$ and $\Omega_\lambda^c = \{2b+1,2b+3,2b+5,2b+7\}$, whereas for the sub-leading order the $b$-dependence cancels again and it is $\Omega_\lambda^\mathbb{1} = \{0,2,4\}$, $\Omega_\lambda^s = \{1,3,5\}$ and $\Omega_\lambda^c = \{0,2,4,6\}$.
In order to bring integrands with different powers to the same form, we perform partial integrations as in \cref{eq:partial_int} until all integrals are reduced to the same power $\lambda = 1+2b$.
With $b$ not necessarily an integer, the integrals generally result in hypergeometric functions as in \cref{eq:si(nmz),eq:ci(nmz)}.
We then rewrite the integrals as 
\begin{align}
    \int_0^{\xi_Y} dx\, x^{-n} 
    \begin{rcases}
        \begin{dcases}
            \sin(m x) \\
            \cos(m x)
        \end{dcases}
    \end{rcases}
    & = \int_0^{\infty} dx\, x^{-n} 
    \begin{rcases}
        \begin{dcases}
            \sin(m x) \\
            \cos(m x)
        \end{dcases}
    \end{rcases}
    - \int_{\xi_Y}^\infty dx\, x^{-n} 
    \begin{rcases}
        \begin{dcases}
            \sin(m x) \\
            \cos(m x)
        \end{dcases}
    \end{rcases} \\
    & = \lim_{z\rightarrow 0}
    \begin{rcases}
        \begin{dcases}
            \text{si}(n,m,z) \\
            \text{ci}(n,m,z)
        \end{dcases}
    \end{rcases}
    -
    \begin{rcases}
        \begin{dcases}
            \text{si}(n,m,\xi_Y) \\
            \text{ci}(n,m,\xi_Y)
        \end{dcases}
    \end{rcases} \,,
    \label{eq:split_integral}
\end{align}
where the first term may diverge at the origin, if considered individually.
We check the exact cancellation of such divergences explicitly by expanding the generalized sine and cosine integrals \cref{eq:si(nmz),eq:ci(nmz)} at the origin, yielding
\begin{align}
    \lim_{z\rightarrow 0}
    \begin{rcases}
        \begin{dcases}
            \text{si}(n,m,z) \\
            \text{ci}(n,m,z)
        \end{dcases}
    \end{rcases}
    \propto
    \begin{rcases}
        \begin{dcases}
            \text{const.}+ z^{2-n} \\
            \text{const.}+ z^{1-n}
        \end{dcases}
    \end{rcases} \,,
\end{align}
as the hypergeometric function goes to 1 at $z=0$.
Even though for large enough $n$ these terms diverge, in the sum of all contributions to $\mathcal{I}^Y(\tilde{x}\ll 1,u,v)$ the divergent terms do indeed cancel exactly, such that only the finite terms stemming from the hypergeometric functions evaluated at $z=\xi_Y$ survive.

\subsection{Subhorizon integrals}
After applying the trigonometric identities \cref{eq:trig_identities} to the subhorizon integrals of \cref{eq:KernelI_split}, the integrals schematically take the form
\begin{align}
    \mathcal{I}_J^{\tilde{x}\gg 1}(u,v) = \int_0^{\xi_J} d\tilde{x}\, \sum_{\tilde{\vartheta} \in \Omega_{\tilde{\vartheta}}} \Bigg( & \sum_{\lambda \in \Omega_\lambda^s } \tilde{c}^s_{\lambda,\tilde{\vartheta}} \tilde{x}^{-\lambda} \sin((c_s \tilde{\vartheta}-1) \tilde{x}+\frac{b\pi}{2}) \nonumber \\
    + & \sum_{\lambda \in \Omega_\lambda^c } \tilde{c}^c_{\lambda,\tilde{\vartheta}} \tilde{x}^{-\lambda} \cos((c_s \tilde{\vartheta}-1) \tilde{x}+\frac{b\pi}{2}) \Bigg)
    \label{eq:KernelIJ_sub_integral}
\end{align}
where $\Omega_{\tilde{\vartheta}} = \{0,u,v,-u,-v,u+v,u-v,-u+v,-u-v\}$, $\Omega_\lambda^s = \{b+1,b+3,b+5,b+7\}$ and $\Omega_\lambda^c = \{b+2,b+4,b+6\}$.
Again, the prefactors are independent of $\tilde{x}$.
Let us note that due to the trigonometric addition rules we used, the coefficients of the (cosine) sine terms are (anti-)symmetric in $\tilde{\vartheta}$, i.e.~$\tilde{c}_{\lambda,\tilde{\vartheta}}^c = -\tilde{c}_{\lambda,-\tilde{\vartheta}}^c$ and $\tilde{c}_{\lambda,\tilde{\vartheta}}^s = \tilde{c}_{\lambda,-\tilde{\vartheta}}^s$.
Next, we perform a number of partial integrations as in \cref{eq:partial_int} on $\mathcal{I}_J^{\tilde{x}\gg 1}(u,v)$, in order to reduce all powers of $\tilde{x}$ multiplying a sine or cosine in the integrals to $\lambda=1+b$.
The resulting integrals can then be evaluated in terms of the generalized sine and cosine integrals \cref{eq:si_D(nmzD),eq:ci_D(nmzD)}.
For $\mathcal{I}_Y^{\tilde{x}\gg 1}(u,v)$ we follow the same strategy, and in this case it is $\Omega_\lambda^s = \{b+2,b+4,b+6\}$ and $\Omega_\lambda^c = \{b+1,b+3,b+5,b+7\}$.

\subsection{Superhorizon kernel for Dirac delta spectrum} \label{app:scaling_I_JY_superhorizon}
To better understand the behavior of $\Omega_{\rm GW}(k)$ in the IR regime and for soft EoS, let us examine the superhorizon kernels $\mathcal{I}_{J/Y}^{\tilde{x}\ll 1}$ for the Dirac delta case more closely. Recall that in this case it is $u=v=k_p/k$ and $v \in [1/2,\infty]$.
Neglecting prefactors and focusing on $b>0$, the integrals schematically behave like
\begin{align}
    \mathcal{I}_J^{\tilde{x}\ll1} (v) \sim \int_0^{\xi_J} d\tilde{x} \,\tilde{x}^{2b+2} \tilde{f}(v\tilde{x})  \qquad \text{and} \qquad
    \mathcal{I}_Y^{\tilde{x}\ll1} (v) \sim \int_0^{\xi_Y} d\tilde{x} \,\tilde{x} \tilde{f}(v\tilde{x}) \,.
\end{align}
Defining $z=v\tilde{x}$ we have
\begin{align}
    \mathcal{I}_J^{\tilde{x}\ll1} (v) \sim v^{-3-2b} \int_0^{v \xi_J} dz \,z^{2b+2} \tilde{f}(z)  \qquad \text{and} \qquad
    \mathcal{I}_Y^{\tilde{x}\ll1} (v) \sim v^{-2} \int_0^{v \xi_Y} dz \,z \tilde{f}(z) \,.
\end{align}
The prefactors decay for large $v$, whereas the integrals grow upon increasing $v$ as long as $\tilde{f}(z)>0$. This indicates that one can expect the peak of the spectrum around $v=k_p/k\sim 1$, which is indeed what we observe.
To get an estimate of the scaling with $v$, we can insert the asymptotic behavior of $\Phi_{\rm iso}$ given in \cref{eq:Phi_iso_scaling}, and split the integrals into $z<1$ and $z>1$ parts.
For $v\sim 1$, we find that $\mathcal{I}_J^{\tilde{x}\ll1} (v) \sim \mathcal{I}_J^{\tilde{x}\ll1} (v)\sim \rm{const.}$.
For $v\gg 1$, on the other hand, we find
\begin{align}
    \mathcal{I}_J^{\tilde{x}\ll1} (v) \sim v^{-5} + v^{-4} \qquad \text{and} \qquad
    \mathcal{I}_Y^{\tilde{x}\ll1} (v) \sim v^{-4} + v^{-4+2b} \,.
    \label{eq:I_JY_peak_scaling}
\end{align}
By plotting the exact results \cref{eq:KernelIJ_super,eq:KernelIY_super} with $u=v$, we find that $\mathcal{I}_J^{\tilde{x}\ll1} (v)$ scales with $v^{-4}$ in the entire range, whereas $\mathcal{I}_Y^{\tilde{x}\ll1} (v)$ scales as $v^{-4}$ near $v\sim 1$ and changes slope to $v^{-4+2b}$ for $v\gg 1$. The $v^{-5}$ and $v^0$ contributions are found to be negligible. Further we note that near $v\sim 1$, $\mathcal{I}_J^{\tilde{x}\ll1}$ is larger, whereas at large $v\gg 1$, $\mathcal{I}_Y^{\tilde{x}\ll1}$ becomes dominant.
For $b<0$, we observe that both $\mathcal{I}_J^{\tilde{x}\ll1} (v)$ and $\mathcal{I}_Y^{\tilde{x}\ll1} (v)$ scale as $v^{-4}$ in the entire range.

\begin{figure}
\centering
\includegraphics[width=0.6\textwidth]{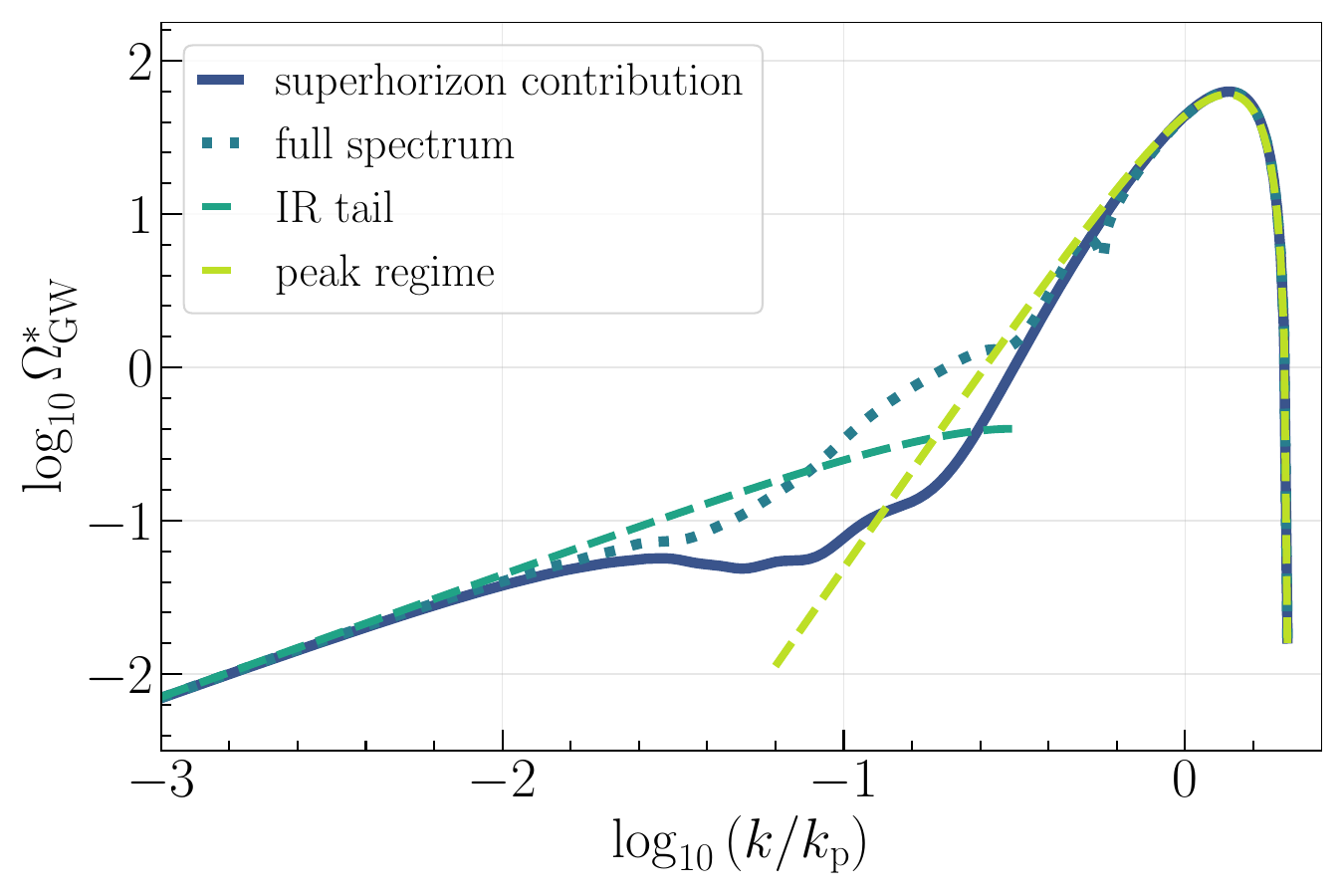}
\caption{Here we show the superhorizon contribution to the GW spectrum \cref{eq:Omega_GW_norm} (solid line) for a soft EoS with $b=3/5$, together with the analytical expression for the IR tail given in \cref{eq:IRscalingomega} and the scaling derived for the peak regime in \cref{eq:Omega_GW_peak_soft} (dashed lines). We fixed the amplitude for \cref{eq:Omega_GW_peak_soft} at $k/k_p=1$ by hand to show the good agreement of the $k$-dependence. For comparison, we also show the full GW spectrum including the subhorizon terms (dotted line).}
\label{fig:Omega_GW_sub_super_scaling}
\end{figure}

The behaviour of the kernel for large $v\gg 1$ results in the scaling of the IR tail of the induced GW spectrum given in \cref{eq:IRscalingomega}.
From the observed behavior for $v\sim 1$ and the observation in \cref{fig:OmegaGW_sub_super}, that the superhorizon contribution dominates the spectrum in the peak regime for very soft EoS, we conclude that the induced GW spectrum for $b\gtrsim 1/3$ roughly scales as
\begin{equation}
    \Omega_{\rm GW}^*(\tilde{k}\lesssim 1) \sim \left(1-\frac{\tilde{k}^2}{4}\right)^2 \, \tilde{k}^{2+2b}\, , \label{eq:Omega_GW_peak_soft}
\end{equation}
near the peak with $\tilde{k}=k/k_p \lesssim 1$.
We show the IR tail and near-peak behaviour, \cref{eq:IRscalingomega,eq:Omega_GW_peak_soft}, respectively, together with the superhorizon contribution and the full spectrum for $b=3/5$ in \cref{fig:Omega_GW_sub_super_scaling}.

%
%

\section{Explicit expressions for coefficients}\label{app:coefficients}
In this appendix, we provide the full expressions for the coefficients appearing in the kernel $I(x,k,u,v)$ as defined by \cref{eq:KernelI_split,eq:KernelI_sub_super_split}. All coefficients presented in the following share the common prefactor
\begin{equation}
    \alpha = \frac{9 \pi ^3 2^{b-4} (b+1)^{2 b+2}}{u^5 v^5 w^3} \,.
\end{equation}

\subsection{Superhorizon contributions}\label{app:coefficients_super}
Due to the symmetry in the momenta $(u,v)$, some of the coefficients are related as
\begin{equation}
    C^{i}_{v} = C^{i}_{u}(u\longleftrightarrow v) \quad \text{and} \quad C^{i}_{u-v} = - C^{i}_{u+v}(v\rightarrow -v) \,,
\end{equation}
and in the same way for the respective $\mathcal{C}_\vartheta^i$.
The remaining independent coefficients appearing in the expression for $\mathcal{I}_J(\tilde{x}\ll 1,u,v)$ are given by
\begin{align}
    C^\mathbb{1} = & \frac{2^{-b-\frac{5}{2}} u v w \alpha}{3 \pi ^2 (b+1) (b+2) \xi_J^3 \Gamma \left(b+\frac{5}{2}\right)} \Big(-8 (2 b+3) (b (b+3)+6) \nonumber \\
    & -12 \xi_J^2 \left(-b^2+(b+1) (b+2) (2 b+3) w \left(u^2+v^2\right)-3 b-6\right) \nonumber \\
    & + 6 \xi_J^4 (b+1) (b+2) w  \left((2 b+3) u^2 v^2 w-(u^2+v^2)\right) -\xi_J^6(b+1) (b+2) u^2 v^2 w^2 \Big)\,,
    \label{eq:coeff_C1}
\end{align}
\begin{align}
    C^{s}_{u} = & \frac{2^{-b-\frac{11}{2}} v \sqrt{w} \alpha}{3 \pi ^2 (b-1) (b+1) (b+2) (b+4) \xi_J^4 \Gamma \left(b+\frac{5}{2}\right)} \Big(270 b (b+3) (2 b+3) (b (b+3)+8) \nonumber \\
    & -24 w \xi_J^4 \Big(u^2 \left(-7 b^2+4 (b-1) (b+1) (b+4) (2 b+3) v^2 w-21 b-32\right) \nonumber \\
    & +(b+1) (b+2) (11 b (b+3)+16) v^2\Big)+96 (b-1) (b+1) (b+4) (2 b+3) u^2 v^2 w^2 \xi_J^5 \nonumber \\
    & +\xi_J^2 \Big((2 b+3) w \left((b (b+3) (19 b (b+3)-592)-1536) u^2+270 b (b+1) (b+2) (b+3) v^2\right) \nonumber \\
    &-270 b (b+3) (b (b+3)+8)\Big)\Big)\,,
\end{align}
\begin{align}
    C^{c}_{u} = & \frac{2^{-b-\frac{11}{2}} u v w \alpha}{3 \pi ^2 (b-1) (b+1) (b+2) (b+4) \xi_J^3 \Gamma \left(b+\frac{5}{2}\right)} \Big( 6 (b+1) (b+2) (7 b (b+3)+32) v^2 w \xi_J^4\nonumber \\
    & +\xi_J^2 \Big((2 b+3) w \left(b (b+3) (19 b (b+3)-256) u^2-6 (b+1) (b+2) (29 b (b+3)+64) v^2\right) \nonumber \\
    & +6 b (b+3) (29 b (b+3)+328)+2304\Big) -2 (2 b+3) (b (b+3) (103 b (b+3)+1016)+768)\Big)\,,
\end{align}
\begin{align}
    C^{Si}_{u} = & \frac{2^{-b-\frac{11}{2}} b (b+3) v w^{3/2} \alpha}{3 \pi ^2 (b-1) (b+1) (b+2) (b+4) \Gamma \left(b+\frac{5}{2}\right)} \Big(90 \left((b (b+3)+4) u^2+3 (b+1) (b+2) v^2\right) \nonumber \\
    &+(2 b+3) u^2 w \left((19 b (b+3)-256) u^2-90 (b+1) (b+2) v^2\right)\Big)\,,
\end{align}
\begin{align}
    C^{s}_{u+v} = & \frac{2^{-b-\frac{15}{2}} \sqrt{w} \alpha}{3 \pi ^2 (b-1)^2 (b+1) (b+2) (b+4)^2 \xi_J^4 \Gamma \left(b+\frac{5}{2}\right) (u+v)^3}\Big(
    \big(3780 b^7+39690 b^6+202230 b^5 \nonumber \\
    &+623700 b^4+1072170 b^3+818910 b^2+155520 b\big) (u^4+v^4)+\big(15120 b^7+158760 b^6 \nonumber \\
    &+808920 b^5+2494800 b^4+4288680 b^3+3275640 b^2+622080 b\big) (u^3 v+u v^3) +\big(22680 b^7 \nonumber \\
    & +238140 b^6 +1213380 b^5+3742200 b^4+6433020 b^3+4913460 b^2+933120 b\big) u^2 v^2 \nonumber \\
    &+\big(-3 (7 b (b+3)+32)^2 \xi_J^6 +6 (2 b+3) (7 b (b+3)+32)^2 \xi_J^5 \big)u^3 v^3 (u+v)^2 w^2-\xi_J^2 (u+v)^4 \nonumber \\
    & \times \Big(135 b (b+3) (b (b+3) (11 b (b+3)+149)+128)+2 (2 b+3) w \Big((b (b+3)(b (b+3)  \nonumber \\
    &\times (19 b (b+3)+3787)+7312)+6144) u v -540 (b-1) b (b+3) (b+4) (u^2+v^2)\Big)\Big) \nonumber \\
    &+3 u v w \xi_J^4 \Big((7 b (b+3)+32) (119 b (b+3)+64) (u^4+v^4)\nonumber \\
    &+(b (b+3) (b (b+3) (49 b (b+3)+4456)+17920)+12288) (u^3 v+u v^3) \nonumber \\
    &+2 (b (b+3) (b (b+3) (49 b (b+3)+3672)+14112)+11264) u^2 v^2\Big)\Big)\,,
\end{align}
\begin{align}
    C^{c}_{u+v} = & \frac{2^{-b-\frac{15}{2}} \alpha}{3 \pi ^2 (b-1)^2 (b+1) (b+2) (b+4)^2 \xi_J^5 \Gamma \left(b+\frac{5}{2}\right) (u+v)^2}\Bigg(-\xi_J^2 (u+v)^2 \Big(2025 b^2 (b (b+3) \nonumber \\
    &+11) (b+3)^2+2 (2 b+3) w \Big(135 b (b+3) (29 b (b+3)+64) (u^2+v^2) \nonumber \\
    &+(b (b+3) (b (b+3) (739 b (b+3)+11077)+15232)+6144) u v \Big)\Big) \nonumber \\
    &+2430 b^2 (b+3)^2 (2 b+3) (b (b+3)+11) (u+v)^2 -6 (7 b (b+3)+32) u^2 v^2 w^2 \xi_J^6 \nonumber \\
    & \times\Big((22 b (b+3)+32) (u^2+v^2)+3 (17 b (b+3)+32) u v\Big)+w \xi_J^4 (u+v)^2 \Big(3 \Big(-45 b (b+3) \nonumber \\
    &\times (b (b+3) (4 b (b+3)-29)-128) (u^2+v^2) +(b (b+3) (b (b+3) (379 b (b+3)\nonumber \\
    &+5857)+9472)+6144) u v \Big) -2 (2 b+3) w \Big(-540 (b-1) b (b+3) (b+4) (u^4+v^4)\nonumber \\
    &+(b-1) b (b+3) (b+4) (19 b (b+3)+284) (u^3 v+uv^3) \nonumber \\
    &-(b (b+3) (b (b+3) (109 b (b+3)+952)+3712)+6144) u^2 v^2\Big)\Big)\Bigg) \,,
\end{align}
\begin{align}
    C^{Si}_{u+v} = & -\frac{2^{-b-\frac{13}{2}} b (b+3) w^{3/2} (u+v) \alpha}{3 \pi ^2 (b-1) (b+1) (b+2) (b+4) \Gamma \left(b+\frac{5}{2}\right)} \Big(90 \Big(3 (b+1) (b+2) (u^2+v^2) \nonumber \\
    &-2 (b (b+3)+1) u v\Big)+(2 b+3) w \Big((19 b (b+3)+284) (u^3 v+u v^3) \nonumber \\
    &-(109 b (b+3)+464) u^2 v^2-540 (u^4+v^4)\Big)\Big)\,.
\end{align}

The coefficients of the $\mathcal{I}_Y^{\tilde{x}\ll 1}(u,v)$ expression are
\begin{align}
    \mathcal{C}^\mathbb{1} = & \frac{2^{-b-\frac{1}{2}} u v w \alpha}{3 \pi ^3 b(b+1) (b+2)^2 \xi_Y^4} \Bigg(b(b+2) \xi_Y \sin (\pi  b) \Gamma \left(-b-\frac{1}{2}\right) \Big(-3 (b+1) (b+2) u^2 v^2 w^2 \xi_Y^4 \nonumber \\
    &+6 (b+1) (b+2) w \xi_Y^2 \left(u^2+v^2\right)+4 b (b+3)+24\Big)-3\ 4^b \xi_Y^{-2 b} \Gamma \left(b+\frac{1}{2}\right) \nonumber \\
    & \times \Big((b+1) (b+2)^2 u^2 v^2 w^2 \xi_Y^4+2 b (b+2)^2 w \xi_Y^2 \left(u^2+v^2\right)+4 b (b (b+3)+6)\Big)\Bigg)\,,
    \label{eq:coeff_calC1}
\end{align}
\begin{align}
    \mathcal{C}^s_u = & \frac{2^{-b-\frac{9}{2}} v \sqrt{w} \alpha}{3 \pi ^3 (b-1) (b+1) (b+2)^2 (b+4) \xi_Y^5} \Bigg(-\frac{3\ 2^{2 b+3} \xi_Y^{-2 b} \Gamma\left(b+\frac{1}{2}\right)}{(2 b+1) (2 b+3) (2 b+5)}\Big(2 u^2 w^2 \xi_Y^4 \Big((b+2)^2  (2 b \nonumber \\
    &+5) (b (b (b (7 b+20)+23)+46)+24) v^2  -(b-2) (b (b (b (7 b+34)+71)+128)+120) u^2\Big) \nonumber \\
    & +(2 b+1) w \xi_Y^2 \Big(4 (b (b (b (b (b (7 b+55)+275)+950)+1815)+1758)+720) u^2 \nonumber \\
    & -45 b (b+1) (b+2)^2 (b+3) (2 b+5) v^2\Big)-90 b (b+2) (b+3) (2 b+1) (2 b+3) (b (b+3)+8)\Big) \nonumber \\
    & -(b+2) \xi_Y \sin (\pi  b) \Gamma \left(-b-\frac{1}{2}\right) \Big(96 (b-1) b (b+4) u^2 v^2 w^2 \xi_Y^5-96 (b-1) b (b+4) u^2 v^2 w^2 \xi_Y^4 \nonumber \\
    & +w \xi_Y^2 \left((b (b+3) (19 b (b+3)-592)-1536) u^2+270 b (b+1) (b+2) (b+3) v^2\right) \nonumber \\
    & +270 b (b+3) (b (b+3)+8)\Big)\Bigg)\,,
\end{align}
\begin{align}
    \mathcal{C}^c_u = & \frac{2^{-b-\frac{9}{2}} u v w \alpha}{3 \pi ^3 (b-1) (b+1) (b+2)^2 (b+4) \xi_Y^4} \Bigg(-(b+2) \xi_Y \sin (\pi  b) \Gamma \left(-b-\frac{1}{2}\right) \Big(w \xi_Y^2 \Big(b (b+3)  \nonumber \\
    & \times(19 b (b+3)-256) u^2-6 (b+1) (b+2) (29 b (b+3)+64) v^2\Big)  -2 (b (b+3) (103 b (b+3)\nonumber \\
    &+1016)+768)\Big)-\frac{3\ 2^{2 b+3} \xi_Y^{-2 b} \Gamma \left(b+\frac{1}{2}\right)}{(2 b+3) (2 b+5)}\Big(w \xi_Y^2 \Big(2 (b-2) (b (b (b (7 b+34)+71)+128) \nonumber \\
    & +120) u^2 +(b+2)^2 (2 b+5) (b (b (37 b+144)+131)+48) v^2\Big)+2 (2 b+3) (b (b (b (b (37 b \nonumber \\
    & +292)+1097) +2342)+2232)+480)\Big)\Bigg)\,,
\end{align}
\begin{align}
    \mathcal{C}^{\mathfrak{ci}}_u = & -\frac{2^{b-\frac{1}{2}} u^3 v w^3 \Gamma \left(b+\frac{1}{2}\right) \alpha}{\pi ^3 (b-1) (b+1) (b+2)^2 (b+4) (2 b+1) (2 b+3) (2 b+5)} \Big((b+2)^2 (2 b+5) \big(b (b (b (7 b+20) \nonumber \\
    & +23)+46)+24\big) v^2-(b-2) (b (b (b (7 b+34)+71)+128)+120) u^2\Big)
    \label{eq:coeff_u_ci} \\
    \mathcal{C}^{Si}_u = & -\frac{2^{-b-\frac{9}{2}} b (b+3) u^2 v w^{5/2} \sin (\pi  b) \Gamma \left(-b-\frac{1}{2}\right) \alpha}{3 \pi ^3 (b-1) (b+1) (b+2) (b+4)} \left((19 b (b+3)-256) u^2-90 (b+1) (b+2) v^2\right)\,,
\end{align}
\begin{align}
    \mathcal{C}^s_{u+v} = & \frac{2^{-b-\frac{15}{2}} \sqrt{w} \alpha}{3 \pi ^3 (b-1)^2 (b+1)^2 (b+2)^2 (b+4)^2 (2 b+1) \xi_Y^5 (u+v)}  \Bigg(-8 (b+1) (b+2) \xi_Y \sin (\pi  b) \nonumber \\
    &  \times \Gamma \left(\frac{1}{2}-b\right) \Big(-675 b^2 (b+3)^2 u^3 v^3 w^2 \xi_Y^5+48 (b-1) (b+4) (11 b (b+3)+16) u^3 v^3 w^2 \xi_Y^4 \nonumber \\
    & +w \xi_Y^2 (u+v)^2 \Big(-540 (b-1) b (b+3) (b+4) (u^2+v^2) +(b (b+3) (b (b+3) (19 b (b+3)\nonumber \\
    &  +3787)+7312)+6144) u v\Big) -135 b (b+3) (7 b (b+3) (b (b+3)+13)+64) (u+v)^2\Big) \nonumber \\
    & -\frac{3\ 4^b \xi_Y^{-2 b} \Gamma \left(b+\frac{1}{2}\right) (u+v)^2}{(2 b+3) (2 b+5)} \Big(2 (b+1) (2 b+1) w \xi_Y^2 \Big(-45 b (b+3) ((b-3) b-7) (b (4 b (7 b \nonumber \\
    &+20)+161)+256) (u^2+v^2) +2 \big(b (b (b (b (b (b (148 b (7 b+76)+71821)+315460)+858280)\nonumber \\
    & +1339429)+1113402) +449664)+92160\big) u v\Big)+w^2 \xi_Y^4 \Big(45 b (b+3) ((b-3) b-7) (b (4 b (7 b \nonumber \\
    & +20)+161)+256) (u^4+v^4) -2 (b+1) \big(b (b (b (b (b (4 b (259 b+572)-14035)-69820) \nonumber \\
    &-185440)-317521)-236288) -30720\big) (u^3 v+u v^3)+2 \big(b (b (b (b (b (2 b (b (28 b (7 b-20) \nonumber \\
    &-14543)-75892)-469385)-997502) -1409050)-1197229)-503296)-61440\big) u^2 v^2\Big)   \nonumber \\
    & -180 b (b+1) (b+2) (b+3) (2 b+1) (2 b+3) \big(b (b (b (74 b+399)+1457)+2283)+512\big)\Big)\Bigg)\,,
\end{align}
\begin{align}
    \mathcal{C}^c_{u+v} = & \frac{2^{-b-\frac{15}{2}} \alpha}{3 \pi ^3 (b-1)^2 (b+1)^2 (b+2)^2 (b+4)^2 \xi_Y^6} \Bigg(4 (b+1) (b+2) \xi_Y \sin (\pi  b) \Gamma \left(-b-\frac{1}{2}\right) \nonumber \\
    & \times \Big(-1215 b^2 (b+3)^2 (b (b+3)+11)+w \xi_Y^2 \Big(135 b (b+3) (29 b (b+3)+64) (u^2+v^2) \nonumber \\
    & +(b (b+3) (b (b+3) (739 b (b+3)+11077)+15232)+6144) u v \Big) +w^2 \xi_Y^4 \Big(-540 (b\nonumber \\
    &-1) b (b+3) (b+4) (u^4+v^4) +(b-1) b (b+3) (b+4) (19 b (b+3)+284) (u^3 v+u v^3) \nonumber \\
    & -(b (b+3) (b (b+3) (109 b (b+3)+952)+3712)+6144) u^2 v^2\Big)\Big) +\frac{3\ 4^b \xi_Y^{-2 b} \Gamma \left(b+\frac{1}{2}\right)}{(2 b+3) (2 b+5)}\nonumber \\
    &  \times\Big(8100 b^2 (b+1) (b+2) (b+3) (2 b+3) (2 b+5) (b (b+3)+11) -2 (b+1) (2 b+3) w \xi_Y^2  \nonumber \\
    & \times\Big(45 b (b+3) (b (b (b (4 b (7 b+41)+985)+3089)+4081)+2048) (u^2+v^2) \nonumber \\
    & +2 \big(b (b (b (b (b (37 b (74 b+761)+146120)+466850)+852185)+766682)+258048) \nonumber \\
    & +30720\big) u v\Big) +w^2 \xi_Y^4 \Big(45 b (b+3) ((b-3) b-7) (b (4 b (7 b+20)+161)+256) (u^4+v^4) \nonumber \\
    & -2 (b+1) \big(b (b (b (b (b (4 b (259 b+572)-14035)-69820)-185440)-317521)-236288) \nonumber \\
    & -30720\big) (u^3 v+u v^3)+2 \big(b (b (b (b (b (2 b (b (28 b (7 b+68)+17489)+108356)+757735) \nonumber \\
    & +1579378)+2048822)+1654707)+795648)+184320\big) u^2 v^2 \Big)\Big)\Bigg)\,,
\end{align}
\begin{align}
   \mathcal{C}^{\mathfrak{ci}}_{u+v} = & -\frac{2^{b-\frac{15}{2}} w^3 \Gamma \left(b+\frac{1}{2}\right) \alpha}{\pi ^3 (b-1)^2 (b+1)^2 (b+2)^2 (b+4)^2 (2 b+1) (2 b+3) (2 b+5)} \Big(45 b (b+3) ((b-3) b-7)  \nonumber \\
    & \times(b (4 b (7 b+20)+161)+256) (u^6+v^6)+2 \big(b (b (b (b (b (b (15347-4 b (259 b+516)) \nonumber \\
    & +70940)+182720)+311441)+217344)+25088)+30720\big) (u^5 v+u v^5) \nonumber \\
    &+b \big(b (b (b (b (4 b (4 b (7 b (7 b-57) -4388)-63245)-616265)-1046504)-997776) \nonumber \\
    &-515687)-180480\big) (u^4 v^2+u^2 v^4) +4 \big(b (b (b (b (b (b (14 b (14 b (6 b+43)+2411)+110157) \nonumber \\
    & +240120)+321144)+308097)+238976)+124928)+30720\big) u^3 v^3 \Big)\,,
    \label{eq:coeff_u+v_ci}
\end{align}
\begin{align}
    \mathcal{C}^{Si}_{u+v} = & \frac{2^{-b-\frac{11}{2}} b (b+3) w^{5/2} \sin (\pi  b) \Gamma \left(-b-\frac{1}{2}\right) (u+v) \alpha}{3 \pi ^3 (b-1) (b+1) (b+2) (b+4)} \Big((19 b (b+3)+284) (u^3 v+u v^3) \nonumber \\
    & -(109 b (b+3)+464) u^2 v^2-540 (u^4+ v^4)\Big)\,.
\end{align}

\subsection{Subhorizon contributions}\label{app:coefficients_sub}
For the $\tilde{x}\gg 1$ terms we have the relations
\begin{align}
    \tilde{C}^{i}_{v} &= \tilde{C}^{i}_{u}(u\longleftrightarrow v) \,, \quad  \tilde{C}^{i}_{-u} = - \tilde{C}^{i}_{u}(u\rightarrow -u) \,, \quad \tilde{C}^{i}_{-v} = - \tilde{C}^{i}_{v}(v\rightarrow -v) \,, \\
    \tilde{C}^{i}_{u-v} &= - \tilde{C}^{i}_{u+v}(v\rightarrow -v) \,, \quad \tilde{C}^{i}_{-u+v} = - \tilde{C}^{i}_{u+v}(u\rightarrow -u) \,, \quad \tilde{C}^{i}_{-u-v} = \tilde{C}^{i}_{u+v}(v\rightarrow -v, u\rightarrow -u) \,,\nonumber
\end{align}
and in the same way for the respective $\tilde{\mathcal{C}}_{\tilde{\vartheta}}^i$.

For $\mathcal{I}_J^{\tilde{x}\gg 1}(u,v)$ we find the coefficients
\begin{align}
    \tilde{C}^{s}_{0} = & \frac{2 \sqrt{2} u v w \xi_J^{-b-4} \alpha}{\pi ^{5/2} (b+1) (b+2)^2 (b+3) (b+4)}\Big(-2 (b+2) (b+3) (b (b+3)+6) \nonumber \\
    & -\xi_J^2 \left((b+1) (b+2) (b+3) (b+4) w \left(u^2+v^2\right)-2 (b (b+3)+6)\right)\Big) \\
    \tilde{C}^{c}_{0} = & \frac{2 \sqrt{2} u v w \xi_J^{-b-3} \alpha}{\pi ^{5/2} (b+1)^2 (b+2)^2 (b+3) (b+4)} \Big(2 (b+1) (b+2) (b (b+3)+6) \nonumber \\
    & +\xi_J^2 \left((b+1) (b+2) (b+3) (b+4) w \left(u^2+v^2\right)-2 (b (b+3)+6)\right)\Big) \\
    \tilde{C}^{si}_{0} = & \frac{\sqrt{2} u v w \alpha}{\pi ^{5/2} (b+1)^2 (b+2)^2 (b+3) (b+4)} \Big( -(b+1)^2 (b+2)^2 (b+3) (b+4) u^2 v^2 w^2 \nonumber \\
    & +2 (b+1) (b+2) (b+3) (b+4) w \left(u^2+v^2\right)-4 (b (b+3)+6)\Big)\,,
\end{align}
\begin{align}
    \tilde{C}^{s}_{u} = & \frac{v \sqrt{w} \xi_J^{-b-4} \alpha}{4 \sqrt{2} \pi ^{5/2} (b-1) (b+1) (b+2)^2 (b+3) (b+4)^2 (b+5)}\Bigg(\xi_J^2 \Big(-\Big((b+2) (b+3) (b+4) u w^{3/2}  \nonumber \\
    &  \times \left(2 (b-1) (b (7 b-2)+40) u^2+(b+1) (b+5) (b (37 b+66)+32) v^2\right)\Big) \nonumber \\
    & +w \Big(2 (b (b (b (b (b (7 b+31)+36)+251)+1211)+2304)+1920) u^2 \nonumber \\
    &  -45 b (b+1) (b+2) (b+3) (b+4) (b+5) v^2\Big)+2 \big(b (b (b (b (37 b+272)+977)+2122)+2112) \nonumber \\
    &  +960\big) u \sqrt{w}+90 b (b+3) (b (b+3)+8)\Big)+2 (b+2) (b+3) \Big(-(b+4) \big(b (b (b (37 b+214) \nonumber \\
    & +661)+1008)+240\big) u \sqrt{w}-45 b (b+3) (b (b+3)+8)\Big)\Bigg)\,,
\end{align}
\begin{align}
    \tilde{C}^{c}_{u} = & \frac{v \sqrt{w} \xi_J^{-b-5} \alpha}{4 \sqrt{2} \pi ^{5/2} (b-1) (b+1)^2 (b+2)^2 (b+3) (b+4)^2 (b+5)} \Bigg(\xi_J^4 \Big((b+1) u w^{3/2}  \nonumber \\
    & \times \Big(2 (b (b (b (b (14 b+71)+108)+415)+1312)+960) u^2+(b+2) (b+3) (b+4) (b+5) \nonumber \\
    & \times (b (37 b+21)+32) v^2\Big)+(b+2) (b+3) (b+4) u^2 w^2 \Big((b+1) (b+5) \big(b (b (7 b-2)+8) \nonumber \\
    & +32\big) v^2-2 (b-1) (b (7 b-2)+40) u^2\Big)+w \Big(45 b (b+1) (b+2) (b+3) (b+4) (b+5) v^2 \nonumber \\
    & -2 (b (b (b (b (b (7 b-6)-236)-726)-911)+192)+960) u^2\Big)-2 \big(b (b (b (b (37 b+227)+707) \nonumber \\
    & +1357)+1032)+960\big) u \sqrt{w}-90 b (b+3) (b (b+3)+8)\Big)+(b+1) (b+2) \xi_J^2  \nonumber \\
    &  \times\Big((b+3) (b+4) w \left(2 (b (b (b (7 b+19)+136)+318)+240) u^2-45 b (b+1) (b+2) (b+5) v^2\right) \nonumber \\
    & +2 (b (b (b (b (37 b+317)+1247)+2887)+3192)+960) u \sqrt{w}+90 b (b+3) (b (b+3)+8)\Big) \nonumber \\
    & -90 b (b+1) (b+2) (b+3)^2 (b+4) (b (b+3)+8)\Bigg)\,,
\end{align}
\begin{align}
    \tilde{C}^{si}_{u} = & \frac{v \sqrt{w} \alpha}{4 \sqrt{2} \pi ^{5/2} (b-1) (b+1)^2 (b+2)^2 (b+3) (b+4)^2 (b+5)} \Bigg(u w^{3/2} \Big(2 \big(b (b (b (b (b (21 b+79)-57) \nonumber \\
    & -203)+816)+2464)+1920\big) u^2+(b+1) (b+2) (b+3) (b+4) (b+5) (b (37 b-24)+32) v^2\Big) \nonumber \\
    & +b u^2 w^2 \Big((b+1) (b+2) (b+3) (b+4) (b+5) (b (7 b-39)-13) v^2-2 \big(b (b (b (b (21 b+139) \nonumber \\
    & +322)+795)+2243)+2240\big) u^2\Big)+b w \Big(2 (b (b (b (b (43-7 b)+463)+1433)+2268)+840) u^2 \nonumber \\
    & +45 (b+1) (b+2) (b+3) (b+4) (b+5) v^2\Big)-(b+2) (b+3) (b+4) u^3 w^{5/2} \nonumber \\
    & \times \left((b+1) (b+5) (b (b (7 b-2)+8)+32) v^2-2 (b-1) (b (7 b-2)+40) u^2\right) \nonumber \\
    & -2 (b (b (b (b (37 b+182)+437)+592)-48)+960) u \sqrt{w}-90 b (b+3) (b (b+3)+8)\Bigg)\,,
\end{align}
\begin{align}
    \tilde{C}^{s}_{u+v} = & \frac{\xi_J^{-b-6} \alpha}{32 \sqrt{2} \pi ^{5/2} (b+1) (b+3) (b+4)^3 (b+5) (b+6) \left(b^2+b-2\right)^2} \Bigg(-(b+2) (b+3) \xi_J^2  \nonumber \\
    & \times \Big(2025 b^2 (b (b+3)+11) (b+3)^2+w \Big(45 b (b+3) (b (b (b (b (b (7 b+82)+514)+2587) \nonumber \\
    & +8095)+11731)+6144) (u^2+v^2)+(b+4) (b (b (b (b (b (37 b (37 b+502)+112635)\nonumber \\
    & +415235)+923060) +1025031)+374016)+46080) u v \Big)+45 b (b+4) (b (b (b (37 b+206)\nonumber \\
    & +764)+1309)+384) (b+3) \sqrt{w} (u+v)\Big)+\xi_J^4 \Big((b+3) w^2 \Big(45 b \left(b^2-7\right) (b (b (b (7 b+61)\nonumber \\
    & +185)+419)+768) (u^4+v^4) -(b+2) (b (b (b (b (b (b (259 b+2089) +3920)-2600)-28675)\nonumber \\
    & -133649)-199424)-30720) (u^3 v+ u v^3)+(b (b (b (b (b (b (b (7 b (7 b+94)+6536)+55132) \nonumber \\
    & +288155)+878236)+1606972)+1781094)+1148928)+368640) u^2 v^2\Big)+2025 b^2 (b+3)^2 \nonumber \\
    & \times (b (b+3)+11)-\Big(w^{3/2} (u+v) \Big(45 b (b+2) (b+3)(b (b (2 b (b (7 b+39)+70)+711)+1781)  \nonumber \\
    & -384) (u^2+v^2)-(b+4) (b (b (b (b (b (b (37 b (7 b+66) +12803)+62795)+243360)+561631)\nonumber \\
    & +649638)+318912)+92160) u v \Big)\Big)+w \Big(45 b (b+3) (b (b (b (b (b (7 b+8)-149)-319)+265) \nonumber \\
    & +1976)+3072) (u^2+v^2)+(b (b (b (b (b (b (37 b (37 b+470)+107281)+425225)+1094680) \nonumber \\
    & +1725221)+1563810)  +712704)+184320) u v\Big)+45 b (b+3) (b (b (b (b (37 b+264)+1048)\nonumber \\
    &+2565)+2650)+1536) \sqrt{w} (u+v)\Big) +2025 b^2 (b+2) (b+3)^3 (b+4) (b+5) (b (b+3)+11)\Bigg)\,,
\end{align}
\begin{align}
    \tilde{C}^{c}_{u+v} = & \frac{\xi_J^{-b-5} \alpha}{32 \sqrt{2} (b-1)^2 (b+1)^2 (b+2)^2 (b+3) (b+4)^3 (b+5) (b+6) \pi ^{5/2}} \Bigg(
    -\Big(1665 b^{11}+39600 b^{10} \nonumber \\
    & +425250 b^9+2749005 b^8 +11907135 b^7+35829270 b^6+74244960 b^5+101295045 b^4 \nonumber \\
    & +83404350 b^3+35047080 b^2+4976640 b\Big)(u+v) \sqrt{w}
    -2025 b^{10}-38475 b^9-330075 b^8 \nonumber \\
    & -1696950 b^7-5736825 b^6-12881025 b^5-18279675 b^4-14543550 b^3-4811400 b^2 \nonumber \\
    & 
    +\Big((u+v) w^{3/2}\Big(135 b (b+1) (b+3) (b (b (b (b (7 b+31)+18)+206)+950)+768) (u^2+v^2) \nonumber \\
    & -(b (b (b (b (b (b (b (37 b (7 b+57)+5181)+6726)+69315)+440391)+1170941)+1353654) \nonumber \\
    & +655104)+184320) u v\Big) +(b+3) (u+v) \Big(45 b \left(b^2-7\right) (b (b (b (7 b+61)+185)+419)\nonumber \\
    & +768) (u^4+v^4)-(b+2) (b (b (b (b (b (b (259 b+2089)+3920)-2600)-28675)-133649)\nonumber \\
    & -199424)-30720) (v u^3+ u v^3) +(b (b (b (b (b (b (b (7 b (7 b+50)-548)-12944)-70825)\nonumber \\
    & -268964)-690740)-1034842) -736256)-122880) v^2 u^2 \Big) w^{5/2} +\Big(-45 b (b+3)\nonumber \\
    & \times (b (b (b (b (b (21 b+167)+432)+983)+2676)+245)-6144) (u^4+v^4)+(b+1) \nonumber \\
    & \times (b (b (b (b (b (2 b (b (259 b+2542)+10043)+68195)+239080)+447527)+258086)\nonumber \\
    & +40704) +184320) (v u^3+u v^3)-(b (b (b (b (b (b (b (7 b (b (7 b+41)+402)+42918)+280797)\nonumber \\
    & +922731)+1727378) +1960516)+1446222)+872448)+368640) v^2 u^2\Big) w^2+\Big(-45 b (b\nonumber \\
    & +3) (b (b (b (b (b (7 b-29)-413)-1367)-2300)-674)+1536) (u^2+v^2)-(b (b+1)  \nonumber \\
    & \times(b (b (b (b (37 b (37 b+343)+60840)+198785)+382085)+412086)+297984) +184320) \nonumber \\
    & \times u v\Big) w -2025 b^2 (b+3)^2 (b (b+3)+11)-45 b (b+3) (b (b (b (b (37 b+219)+778)+1665)\nonumber \\
    & +1165) +1536) (u+v) \sqrt{w}\Big) \xi_J^4+(b+1) (b+2) \Big((b+3) (u+v) \Big(-45 b \left(b^2-7\right) (b (b (b (7 b\nonumber \\
    & +61) +185)+419)+768) (u^2+v^2)+(b+4) (b (b (b (b (b (37 b (7 b+82)+18945)+90065)\nonumber \\
    & +264020) +368541)+181056)+46080) v u\Big) w^{3/2}+\Big(45 b (b+3)^2 (b (b (b (b (7 b+24)+88) \nonumber \\
    & +735)+1525)+1536) (u^2+v^2)+(b+4) (b (b (b (b (b (37 b (37 b+412)+84105)+290855)\nonumber \\
    & +598970)+637041)+270336) +46080) v u\Big) w +2025 b^2 (b+3)^2 (b (b+3)+11)\nonumber \\
    & +45 b (b+3) (b (b (b (b (37 b+309)+1318)+3465)+4135) +1536) (u+v) \sqrt{w}\Big) \xi_J^2\Bigg)\,,
\end{align}
\begin{align}
    \tilde{C}^{si}_{u+v} = & \Bigg(b (u+v) w^{3/2}\Big(45 (b+3) (b (b (b (b (b (28 b+85)-266)-695)+1168)+4480)+3840) (u^2\nonumber \\
    & +v^2) +(b (b (b (b (b (b (8879-37 b (7 b+20))+66805)+190310)+140479)-376770)\nonumber \\
    & -643584) -357120) v u\Big) +b (u+v) \Big(45 (b+3) (b (b (b (4 b (b (7 b+57)+142)+975)+2149) \nonumber \\
    & -2688)-11520) (u^4+v^4) -(b (b (b (b (b (b (b (777 b+8986)+41089)+117815)+289120)\nonumber \\
    & +393983)-334106) -1530944)-1125120) (v u^3+u v^3)+2 (b (b (b (b (b (b (b (49 b (b+8) \nonumber \\
    &+1658)+14165)+85570) +220646)+114873)-573273)-1197280)-729600) v^2 u^2 \Big) \nonumber \\
    & \times w^{5/2} -(b+3) \Big(45 b \left(b^2-7\right) (b (b (b (7 b+61)+185)+419)+768) (u^6+v^6)+(b (b (b (b \nonumber \\
    & \times(33155 -b (b (b (259 b+1977)+2608)-7000))+143569)+202752)-54272)+61440) \nonumber \\
    & \times (v u^5+u v^5) +b (b (b (b (b (b (b (7 b (7 b-24)-5447)-26395)-75185)-201574)-332457)\nonumber \\
    & -233383) -119040) (v^2 u^4+u^2 v^4)+(b+1) (b (b (b (b (b (b (7 b (21 b+173)+4589)+24475)\nonumber \\
    & +90440)+112258) +34304)+106496)+122880) v^3 u^3\Big) w^3+\Big(-45 b (b+3) (b (b (b (b (b (42 b\nonumber \\
    & +281)+579)+1655) +6144)+5399)-3840) (u^4+v^4)+(b (b (b (b (b (b (b (b (777 b+5821)\nonumber \\
    & +14421)+50997)+276420) +633438)+476334)+53504)+258048)+368640) (v u^3\nonumber \\
    & +u v^3)-b (b (b (b (b (b (b (b (7 b (7 b-33) +486)+48486)+311355)+884271)+1340156)\nonumber \\
    & +1018854)+337854)+184320) v^2 u^2 \Big) w^2+\Big(-45 b^2 (b+3) (b (b (b (b (7 b-66)-632)\nonumber \\
    & -2145)-3965)-1839) (u^2+v^2) -(b (b (b (b (b (b (37 b (37 b+290)+43831)+130475)\nonumber \\
    & +220960)+239771)+257280)-116736) +184320) v u\Big) w -2025 b^2 (b+3)^2 (b (b+3)+11)\nonumber \\
    & -45 b (b+3) (b (b (b (b (37 b+174)+508)+765)-320) +1536) (u+v) \sqrt{w}\Bigg) \alpha \nonumber \\
    & \times 1\Big/\Bigg(32 \sqrt{2} (b-1)^2 (b+1)^2 (b+2)^2 (b+3) (b+4)^3 (b+5) (b+6) \pi ^{5/2}\Bigg)\,.
\end{align}

And finally for $\mathcal{I}_Y^{\tilde{x}\gg 1}(u,v)$ we have
\begin{align}
    \tilde{\mathcal{C}}^{s}_{0} = & \frac{2 \sqrt{2} u v w \xi_Y^{-b-3} \alpha}{\pi ^{5/2} (b+1)^2 (b+2)^2 (b+3) (b+4)}  \Big( 2 (b+1) (b+2) (b (b+3)+6) \nonumber \\
    & +\xi_Y^2 \left((b+1) (b+2) (b+3) (b+4) w \left(u^2+v^2\right)-2 (b (b+3)+6)\right)\Big)\,,\\
    \tilde{\mathcal{C}}^{c}_{0} = & \frac{2 \sqrt{2} u v w \xi_Y^{-b-4} \alpha}{\pi ^{5/2} (b+1) (b+2)^2 (b+3) (b+4)}  \Big(2 (b+2) (b+3) (b (b+3)+6) \nonumber \\
    & +\xi_Y^2 \left((b+1) (b+2) (b+3) (b+4) w \left(u^2+v^2\right)-2 (b (b+3)+6)\right)\Big)\,,\\
    \tilde{\mathcal{C}}^{si}_{0} = & \frac{\sqrt{2} u v w \alpha}{\pi ^{5/2} (b+1)^2 (b+2)^2 (b+3) (b+4)} \Big((b+1)^2 (b+2)^2 (b+3) (b+4) u^2 v^2 w^2 \nonumber \\
    & -2 (b+1) (b+2) (b+3) (b+4) w \left(u^2+v^2\right)+4 b (b+3)+24\Big)\,,
\end{align}
\begin{align}
    \tilde{\mathcal{C}}^{s}_{u} = & \frac{v \sqrt{w} \xi_Y^{-b-5} \alpha}{4 \sqrt{2} \pi ^{5/2} (b-1) (b+1)^2 (b+2)^2 (b+3) (b+4)^2 (b+5)} \Bigg(\xi_Y^4 \Big((b+1) u w^{3/2} \Big(2 (b (b (b (b (14 b+71) \nonumber \\
    & +108)+415)+1312)+960) u^2+(b+2) (b+3) (b+4) (b+5) (b (37 b+21)+32) v^2\Big) \nonumber \\
    & +(b+2) (b+3) (b+4) u^2 w^2 \Big((b+1) (b+5) (b (b (7 b-2)+8)+32) v^2-2 (b-1)  \nonumber \\
    & \times(b (7 b-2)+40) u^2\Big)+w \Big(45 b (b+1) (b+2) (b+3) (b+4) (b+5) v^2-2 (b (b (b (b (b (7 b-6) \nonumber \\
    & -236)-726)-911)+192)+960) u^2\Big)-2 (b (b (b (b (37 b+227)+707)+1357)+1032) \nonumber \\
    & +960) u \sqrt{w}-90 b (b+3) (b (b+3)+8)\Big)+(b+1) (b+2) \xi_Y^2 \Big((b+3) (b+4) w \Big(2 (b (b (b (7 b \nonumber \\
    & +19)+136)+318)+240) u^2-45 b (b+1) (b+2) (b+5) v^2\Big)+2 (b (b (b (b (37 b+317)+1247) \nonumber \\
    & +2887)+3192)+960) u \sqrt{w}+90 b (b+3) (b (b+3)+8)\Big) \nonumber \\
    & -90 b (b+1) (b+2) (b+3)^2 (b+4) (b (b+3)+8)\Bigg)\,,
\end{align}
\begin{align}
    \tilde{\mathcal{C}}^{c}_{u} = & \frac{v \sqrt{w} \xi_Y^{-b-4} \alpha}{4 \sqrt{2} \pi ^{5/2} (b-1) (b+1) (b+2)^2 (b+3) (b+4)^2 (b+5)} \Bigg(\xi_Y^2 \Big((b+2) (b+3) (b+4) u w^{3/2}  \nonumber \\
    & \times \left(2 (b-1) (b (7 b-2)+40) u^2+(b+1) (b+5) (b (37 b+66)+32) v^2\right) \nonumber \\
    & +w \Big(45 b (b+1) (b+2) (b+3) (b+4) (b+5) v^2-2 (b (b (b (b (b (7 b+31)+36)+251)+1211) \nonumber \\
    & +2304)+1920) u^2\Big)-2 (b (b (b (b (37 b+272)+977)+2122)+2112)+960) u \sqrt{w} \nonumber \\
    & -90 b (b+3) (b (b+3)+8)\Big)+2 (b+2) (b+3) \Big((b+4) (b (b (b (37 b+214)+661)+1008) \nonumber \\
    & +240) u \sqrt{w}+45 b (b+3) (b (b+3)+8)\Big)\Bigg)\,,
\end{align}
\begin{align}
    \tilde{\mathcal{C}}^{si}_{u} = & \frac{v \sqrt{w} \alpha}{4 \sqrt{2} \pi ^{5/2} (b-1) (b+1)^2 (b+2)^2 (b+3) (b+4)^2 (b+5)} \Bigg(-u w^{3/2} \Big(2 (b (b (b (b (b (21 b+79)-57) \nonumber \\
    & -203)+816)+2464)+1920) u^2+(b+1) (b+2) (b+3) (b+4) (b+5) (b (37 b-24)+32) v^2\Big) \nonumber \\
    & +b u^2 w^2 \Big(2 (b (b (b (b (21 b+139)+322)+795)+2243)+2240) u^2-(b+1) (b+2) (b+3) \nonumber \\
    & \times (b+4) (b+5) (b (7 b-39)-13) v^2\Big)-b w \Big(2 (b (b (b (b (43-7 b)+463)+1433)+2268) \nonumber \\
    & +840) u^2+45 (b+1) (b+2) (b+3) (b+4) (b+5) v^2\Big)+(b+2) (b+3) (b+4) u^3 w^{5/2} \Big((b+1) \nonumber \\
    &  (b+5) (b (b (7 b-2)+8)+32) v^2-2 (b-1) (b (7 b-2)+40) u^2\Big)+2 (b (b (b (b (37 b+182)+437) \nonumber \\
    & +592)-48)+960) u \sqrt{w}+90 b (b+3) (b (b+3)+8)\Bigg)\,,
\end{align}
\begin{align}
    \tilde{\mathcal{C}}^{c}_{u+v} = & \frac{\xi_Y^{-b-6} \alpha}{32 \sqrt{2} \pi ^{5/2} (b+1) (b+3) (b+4)^3 (b+5) (b+6) \left(b^2+b-2\right)^2} \Bigg((b+2) (b+3) \xi_Y^2 \nonumber \\
    & \times \Big(2025 b^2 (b (b+3)+11) (b+3)^2+w \Big(45 b (b+3) (b (b (b (b (b (7 b+82)+514)+2587)+8095) \nonumber \\
    & +11731)+6144) (u^2+v^2)+(b+4) (b (b (b (b (b (37 b (37 b+502)+112635)+415235) \nonumber \\
    & +923060)+1025031)+374016)+46080) u v \Big)+45 b (b+4) (b (b (b (37 b+206)+764)+1309)  \nonumber \\
    & +384) (b+3) \sqrt{w}(u+v)\Big)-\xi_Y^4 \Big((b+3) w^2 \Big(45 b \left(b^2-7\right) (b (b (b (7 b+61)+185)+419)  \nonumber \\
    & +768)(u^4+v^4)-(b+2) (b (b (b (b (b (b (259 b+2089) +3920)-2600)-28675)-133649) \nonumber \\
    & -199424)-30720)(u^3 v+u v^3)+(b (b (b (b (b (b (b (7 b (7 b+94)+6536)+55132)+288155) \nonumber \\
    &+878236) +1606972)+1781094)+1148928)+368640) u^2 v^2\Big)+2025 b^2 (b+3)^2 (b (b+3) \nonumber \\
    & +11)-w^{3/2} (u+v) \Big(45 b (b+2) (b+3)(b (b (2 b (b (7 b+39)+70)+711)+1781)-384) \nonumber \\
    &\times (u^2+v^2)-(b+4) (b (b (b (b (b (b (37 b (7 b+66)+12803)+62795)+243360)+561631) \nonumber \\
    & +649638)+318912)+92160) u v\Big)+w \Big(45 b (b+3) (b (b (b (b (b (7 b+8)-149)-319)+265) \nonumber \\
    & +1976)+3072) (u^2+v^2)+(b (b (b (b (b (b (37 b (37 b+470)+107281)+425225)+1094680) \nonumber \\
    & +1725221)+1563810)+712704)+184320) u v\Big)+45 b (b+3) (b (b (b (b (37 b+264)+1048) \nonumber \\
    & +2565)+2650)+1536) \sqrt{w} (u+v)\Big)-2025 b^2 (b+2) (b+3)^3 (b+4)(b+5) (b (b+3)+11)\Bigg)\,,
\end{align}
\begin{align}
    \tilde{\mathcal{C}}^{s}_{u+v} = & \frac{\xi_Y^{-b-5} \alpha}{32 \sqrt{2} (b-1)^2 (b+1)^2 (b+2)^2 (b+3) (b+4)^3 (b+5) (b+6) \pi ^{5/2}} \Bigg(-(u+v)\sqrt{w}\Big(1665 b^{11} \nonumber \\
    & +39600 b^{10}+425250 b^9+2749005 b^8+11907135 b^7+35829270 b^6+74244960 b^5 \nonumber \\
    & +101295045 b^4+83404350 b^3+35047080 b^2+4976640 b\Big)-2025 b^{10}-38475 b^9-330075 b^8 \nonumber \\
    & -1696950 b^7-5736825 b^6-12881025 b^5-18279675 b^4-14543550 b^3-4811400 b^2 \nonumber \\
    & 
    +\Big((u+v) \Big(135 b (b+1) (b+3) (b (b (b (b (7 b+31)+18)+206)+950)+768) (u^2+v^2) \nonumber \\
    & -(b (b (b (b (b (b (b (37 b (7 b+57)+5181)+6726)+69315)+440391)+1170941)+1353654) \nonumber \\
    & +655104)+184320) v u\Big) w^{3/2}+(b+3) (u+v) \Big(45 b \left(b^2-7\right) (b (b (b (7 b+61)+185)+419)\nonumber \\
    & +768) (u^4+v^4)-(b+2) (b (b (b (b (b (b (259 b+2089)+3920)-2600)-28675)-133649) \nonumber \\
    & -199424)-30720) (v u^3+v^3 u)+(b (b (b (b (b (b (b (7 b (7 b+50)-548)-12944)-70825) \nonumber \\
    & -268964)-690740)-1034842)-736256)-122880) v^2 u^2\Big) w^{5/2}\nonumber \\
    & +\Big(-45 b (b+3) (b (b (b (b (b (21 b+167)+432)+983)+2676)+245)-6144) (u^4+v^4) \nonumber \\
    & +(b+1) (b (b (b (b (b (2 b (b (259 b+2542)+10043)+68195)+239080)+447527)+258086) \nonumber \\
    & +40704)+184320) (v u^3+v^3 u)-(b (b (b (b (b (b (b (7 b (b (7 b+41)+402)+42918)+280797) \nonumber \\
    & +922731)+1727378)+1960516)+1446222)+872448)+368640) v^2 u^2 \Big) w^2 \nonumber \\
    & +\Big(-45 b (b+3) (b (b (b (b (b (7 b-29)-413)-1367)-2300)-674)+1536) (u^2+v^2) \nonumber \\
    &-(b (b+1) (b (b (b (b (37 b (37 b+343)+60840)+198785)+382085)+412086)+297984)\nonumber \\
    &+184320) v u\Big) w  -2025 b^2 (b+3)^2 (b (b+3)+11)-45 b (b+3) (b (b (b (b (37 b+219)+778) \nonumber \\
    &+1665)+1165) +1536) (u+v) \sqrt{w}\Big) \xi_Y^4+(b+1) (b+2) \Big((b+3) (u+v) \Big(-45 b \left(b^2-7\right) \nonumber \\
    & \times(b (b (b (7 b+61) +185)+419)+768) (u^2+v^2)+(b+4) (b (b (b (b (b (37 b (7 b+82)+18945) \nonumber \\
    &+90065)+264020) +368541)+181056)+46080) v u\Big) w^{3/2} +\Big(45 b (b+3)^2 (b (b (b (b (7 b \nonumber \\
    &+24)+88)+735)+1525)+1536) (u^2+v^2)+(b+4)(b (b (b (b (b (37 b (37 b+412) +84105)\nonumber \\
    & +290855)+598970)+637041)+270336) +46080) v u\Big) w +2025 b^2 (b+3)^2 (b (b+3)+11) \nonumber \\
    &+45 b (b+3) (b (b (b (b (37 b+309)+1318)+3465)+4135) +1536) (u+v) \sqrt{w}\Big) \xi_Y^2\Bigg)\,,
\end{align}
\begin{align}
    \tilde{\mathcal{C}}^{si}_{u+v} = & \Bigg(-b (u+v) \Big(45 (b+3) (b (b (b (b (b (28 b+85)-266)-695)+1168)+4480)+3840) (u^2+v^2) \nonumber \\
    & +(b (b (b (b (b (b (8879-37 b (7 b+20))+66805)+190310)+140479)-376770)-643584) \nonumber \\
    & -357120) v u \Big) w^{3/2}-b (u+v) \Big(45 (b+3) (b (b (b (4 b (b (7 b+57)+142)+975)+2149)-2688) \nonumber \\
    & -11520) (u^4+v^4)-(b (b (b (b (b (b (b (777 b+8986)+41089)+117815)+289120)+393983) \nonumber \\
    & -334106)-1530944)-1125120) (v u^3+v^3 u)+2 (b (b (b (b (b (b (b (49 b (b+8)+1658)+14165) \nonumber \\
    & +85570)+220646)+114873)-573273)-1197280)-729600) v^2 u^2\Big) w^{5/2}+(b+3) \nonumber \\
    & \times \Big(45 b \left(b^2-7\right) (b (b (b (7 b+61)+185)+419)+768) (u^6+v^6)+(b (b (b (b (33155 \nonumber \\
    &-b (b (b (259 b+1977)+2608)-7000))+143569)+202752)-54272)+61440) (v u^5+v^5 u) \nonumber \\
    &+b (b (b (b (b (b (b (7 b (7 b-24)-5447)-26395)-75185)-201574)-332457)-233383) \nonumber \\
    &-119040) (v^2 u^4+ v^4 u^2)+(b+1) (b (b (b (b (b (b (7 b (21 b+173)+4589)+24475)+90440) \nonumber \\
    &+112258)+34304)+106496)+122880) v^3 u^3\Big) w^3+\Big(45 b (b+3) (b (b (b (b (b (42 b+281) \nonumber \\
    &+579)+1655)+6144)+5399)-3840) (u^4+v^4)-(b (b (b (b (b (b (b (b (777 b+5821)+14421) \nonumber \\
    &+50997)+276420)+633438)+476334)+53504)+258048)+368640) (v u^3+v^3 u) \nonumber \\
    &+b (b (b (b (b (b (b (b (7 b (7 b-33)+486)+48486)+311355)+884271)+1340156)+1018854) \nonumber \\
    &+337854)+184320) v^2 u^2 \Big) w^2+\Big(45 b^2 (b+3) (b (b (b (b (7 b-66)-632)-2145)-3965) \nonumber \\
    &-1839) (u^2+v^2)+(b (b (b (b (b (b (37 b (37 b+290)+43831)+130475)+220960)+239771) \nonumber \\
    &+257280)-116736)+184320) v u\Big) w+2025 b^2 (b+3)^2 (b (b+3)+11)+45 b (b+3)  \nonumber \\
    &\times(b (b (b (b (37 b+174)+508)+765)-320)+1536) (u+v) \sqrt{w}\Bigg) \alpha \nonumber \\
    & \times 1 \Big/\Bigg(32 \sqrt{2} (b-1)^2 (b+1)^2\times (b+2)^2 (b+3) (b+4)^3 (b+5) (b+6) \pi ^{5/2}\Bigg)\,.
\end{align}

\bibliography{references}

\end{document}